# Optical frequency comb technology for ultra-broadband radio-frequency photonics

## Victor Torres-Company

*Microtechnology and Nanoscience Department (MC2), Chalmers University of Technology, SE- 41296 Gothenburg, Sweden*

## Andrew M. Weiner

*School of Electrical and Computer Engineering, Purdue University, West Lafayette, IN- 47907, USA*

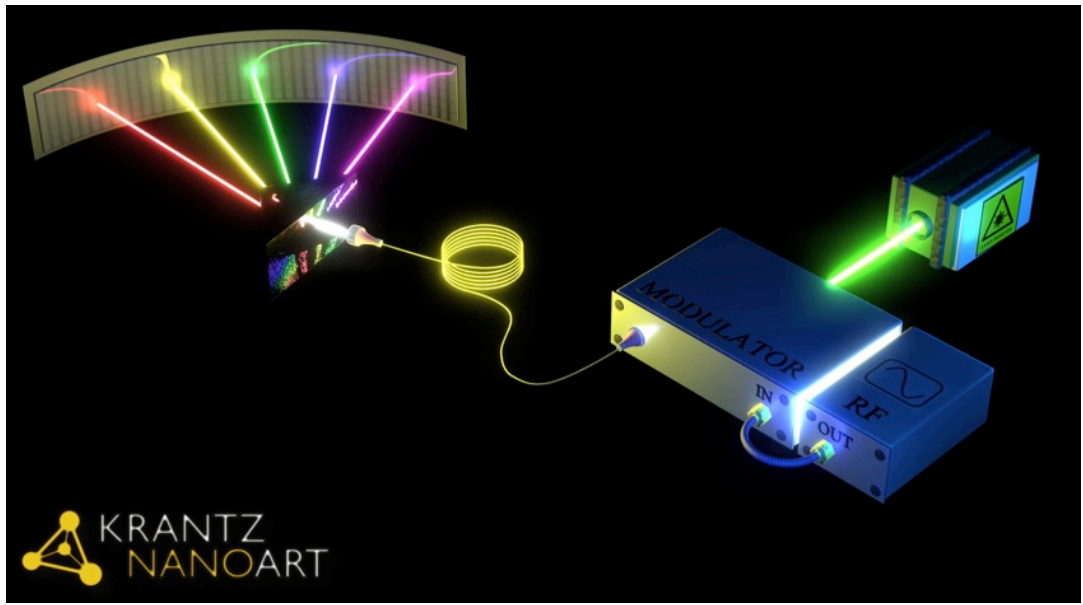

*Abstract.-* The outstanding phase-noise performance of optical frequency combs has led to a revolution in optical synthesis and metrology, covering a myriad of applications, from molecular spectroscopy to laser ranging and optical communications. However, the ideal characteristics of an optical frequency comb are application dependent. In this review, the different techniques for the generation and processing of high-repetition-rate (>10 GHz) optical frequency combs with technologies compatible with optical communication equipment are covered. Particular emphasis is put on the benefits and prospects of this technology in the general field of radio-frequency photonics, including applications in high-performance microwave photonic filtering, ultra-broadband coherent communications, and radio-frequency arbitrary waveform generation.

**Outline**

1. Introduction
2. Opto-electronic frequency comb generators
   a. Monolithic cavity-based optoelectronic frequency comb generators
   b. Flat-top frequency comb generators
   c. From CW to femtosecond pulse generation
   d. Towards higher repetition-rates with parametric comb generators

3. Brief overview on Fourier processing of frequency combs

4. RF signal processing applications
   a. Multi-tap dispersive delay-line RF photonic filters
   b. Phase-programmable filters
   c. Photonic downsampling and reconfigurable filtering

5. Coherent communications and arbitrary waveform generation
   5.1 High-performance microwave synthesis
   5.2 Dynamic radio-frequency waveform generation
   5.3 THz coherent communications
   5.4 Coherent optical communications

6. Discussion

7. Conclusions

8. References

Acknowledgments

# 1. Introduction

The field of radio-frequency (RF) photonics merges photonic technologies with microwave engineering to realize unique applications in telecommunications, radar, sensors, metrology, testing and imaging [1]. The history of RF photonics runs in parallel to the history of optical communications. These fields complement each other and benefit from their respective technological advances. Thanks to the development of the first semiconductor lasers, external modulators and photodetectors offering a bandwidth in the GHz range, together with the first low-loss single-mode fibers, we can view now the 1970's as the birth of RF photonics [2]. Some of the applications based on these technologies (such as antenna remoting, wireless communications or cable television systems) are now commercially established [3].

In a parallel research line, since its demonstration in 2000 [4] [5] [6], the femtosecond optical frequency comb has revolutionized the fields of optical synthesis and metrology [7], and its applications are expanding every day, going from molecular spectroscopy and optical clocks to optical and radio-frequency arbitrary waveform generation [8]. The broad impact of this tool was recognized with half of the 2005 Nobel Prize in Physics awarded to T. W. Hänsch and J. L. Hall.

Yet, the ideal characteristics of an optical frequency comb depend on the target application [8]. For example, applications in optical arbitrary waveform generation [9] are less demanding in terms of bandwidth and absolute stabilization, but require high repetition rate (>10 GHz), robustness, and flexibility in tuning independently the repetition rate and central frequency. These tasks are not easily met by optical frequency combs implemented either with Ti:Sa technology or fiber lasers. High-repetition-rate optical frequency combs with a relatively broad optical bandwidth, compactness, and low noise level offer a huge new avenue of possibilities in radio-frequency photonics.

The goal of this paper is not to provide another general review on microwave photonics. The reader looking for an introduction to the field can find excellent documents in the literature [1] [2] [3]. Our aim is to present the available technologies for the generation and processing of high-repetition-rate optical frequency combs and highlight their unique characteristics for emerging applications in radio-frequency photonics. These include high-performance filtering of microwave signals, ultra-broadband coherent wireless/optical communications, and high-fidelity synthesis of ultra-broadband waveforms.

The reminder of this work is as follows. In Section 2, an opto-electronic platform for frequency comb generation with ~ 10-40 GHz repetition rates is reviewed. The advantages, benefits and the possibility to achieve broader bandwidths and higher repetition rates are also discussed. The state of the art in line-by-line manipation of frequency combs is briefly discussed in Section 3. Section 4 is devoted to the potential of opto-electronic frequency comb generators for the processing of microwave signals, whereas applications of high-performance waveform synthesis in fiber-optic as well as wireless communications are highlighted in Section 5. Recent trends and challenges towards the development of more compact platforms are discussed in Section 6. Finally, the concluding remarks are briefly expressed in Section 7.

## 2. Opto-electronic frequency comb generators

An optical frequency comb consists of a series of evenly spaced discrete spectral components. However, not every multi-wavelength source can qualify as an optical frequency comb. The true potential of this instrument relies on: i/ maintaining high spectral coherence across the whole bandwidth and ii/ the possibility to synthesize in an independent manner and with great accuracy the offset and spacing between lines [8].

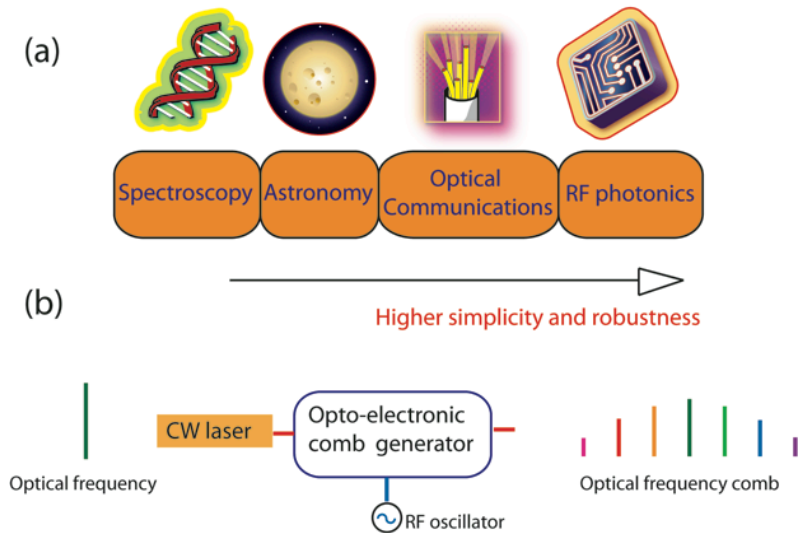

**Fig. 1**. *(a) Different frequency comb applications that benefit from high repetition rate sources. (b) General layout of an opto-electronic frequency comb generator. This platform combines robustness and simplicity and is therefore better suited for applications in RF photonics and optical communications.*

With the advent of self-referenced optical frequency combs in Ti:Sa [4] or Er-fiber lasers [10], researchers have at their disposal a metrology/synthesis tool capable to deliver a performance level that was previously attained only by very specialized laboratories in the world. Femtosecond optical frequency combs provide millions of lines, with reported relative accuracies at the $10^{-19}$ level [11], and often offer an octave-spanning bandwidth spectrum. However, apart from one notable exception [12], self-referenced frequency combs operate below 10-GHz repetition rates. For optical communications and RF photonics applications, self-referencing is not required but, as sketched in Fig. 1(a), higher repetition rates, simplicity, spectral flatness, robustness, and in some cases tunability are needed.

Opto-electronic frequency comb generators can meet these demands [13]. The general scheme is depicted in Fig. 1(b). A continuous-wave (CW) laser is sent to an optical system that contains one or several modulators that are driven by an external RF oscillator. At the output, a comb emerges with the central wavelength defined by the CW laser and the frequency spacing fixed by the oscillator frequency. Note that the name *opto-electronic* comb generator is coined here because in principle any type of modulator could be used for this task. When implemented by electro-optic modulators (as is usually the case), it may be named *electro-optic* comb generator. This configuration has been researched since the 1960's [14], and in the 70's and 80's as a source of high-repetition-rate picosecond pulses [15] [16]. Nowadays, thanks to the advent of low-noise high-frequency RF oscillators [17] and low-driving-voltage high-power handling electro-optic modulators, this platform is gaining increasing popularity due to its robustness, simplicity and performance. Unlike actively mode-locked lasers, it permits independent electrical tuning of the repetition rate and central frequency.

Let us consider a simple electro-optic phase modulator as the optoelectronic black box in Fig.1. The optical phase imparted onto the CW laser is $\pi V / V_\pi cos(2\pi f t)$. Here $f$ is the frequency and $V$ is the voltage amplitude of the driving RF signal, respectively. The parameter $V_\pi$, called the half-wave voltage, is a constant that determines the amount of RF power required to achieve a $\pi$ phase change. It depends in an intricate manner on the electro-optic coefficient of the material and the waveguide structure, and its magnitude varies with the particular RF frequency (usually in an increasing manner with increasing frequency). In lithium niobate, this parameter may range from ~2V to 11V or more at 10 GHz. In order to achieve a broad optical bandwidth (e.g., to approach THz span) the so-called modulation index, $\Delta\theta = \pi V / V_\pi$ , must be maximized. Thus, an ideal electro-optic modulator should not only have a low $V_\pi$ parameter, but it should be capable of sustaining high RF power. Of course, larger bandwidths could be achieved by placing several phase modulators in tandem. In this case, the temporal phases should be properly aligned in time (e.g. with phase shifters in the electrical domain) and the effective modulation index will be increased by a factor of *N*, with *N* being the number of modulators.

Another relevant parameter in optoelectronic comb generation is the frequency spacing, which is limited by the modulator's bandwidth. Commercially available lithium niobate modulators can provide a bandwidth around 40 GHz. There are very promising research efforts that have shown bandwidths approaching 100 GHz by optimizing the waveguide design in lithium niobate [18] or exploiting electro-optic modulation in polymer structures [19]. These polymer modulators have an ultrafast (~10 fs) electro-optic response and electro-optic coefficient about an order of magnitude higher than lithium niobate [20] [21].

## 2.1 Coupled-cavity monolithic electro-optic frequency comb generators

Considering a state-of-the-art electro-optic phase modulator with $V_\pi = 2V$ at 10 GHz and a maximum RF power handling of 1W, one could achieve ~30 comb lines at -10 dB bandwidth. This is not very appealing for some demanding applications, particularly because the power from the CW laser does not distribute in a uniform manner across the bandwidth. In order to increase the modulation efficiency, several resonant schemes have been reported (we shall discuss the spectral flatness issue in the following section). The Fabry-Perot modulator [14] consists of an electro-optic phase modulator placed inside a Fabry-Perot cavity and driven at a frequency commensurate with the optical round trip time. This approach to comb generation received significant attention in the frequency metrology community [22] [23] prior to the advent of self-referenced mode-locked lasers. An integrated implementation based on a lithium niobate waveguide phase modulator with mirror coatings deposited on the facets of the chip is commercially available from a company in Japan [24]. This type of frequency comb can provide terahertz bandwidth spectra and sub-10 fs timing jitter when combined with a narrow-linewidth laser and an active stabilization scheme which locks the input laser frequency and the cavity resonance [25]. In addition to applications in optical metrology, such optical frequency comb generators have proved to be useful for high-repetition-rate pulse train generation [26] and for injection locking of widely separated lasers [27]. We may understand the synchronous phase modulation as sweeping the transmission resonance through the CW laser frequency twice per modulation period. This carves the continuous input into two narrow pulses transmitted per period, one red-shifted and the other blue-shifted. The resulting optical power spectrum exhibits a double-sided exponential decay (linear ramp in dB) [26].

Different fiber ring cavities containing a phase modulator or frequency shifter have been also considered in the literature [28]. As in the Fabry-Perot case, the frequency of the oscillator must match precisely the free spectral range (or a multiple integer), the cavity must remain stable in length over long temporal spans to keep the spectral phase coherence and, more importantly, the CW laser must be precisely tuned to one of the optical resonances. In practice, this means that some feedback mechanism is needed to compensate for environmental instabilities, as is done in [25]. Comb generator schemes that overlook these aspects raise questions about the long-term stability of the spectrum. It is worth noting recent efforts to realize highly integrated optoelectronic comb generators in semiconductor materials to alleviate some of these problems [29].

Nevertheless, cavity-based optoelectronic frequency comb generators have a limited range in which the CW and the RF source can be tuned, and the achieved spectrum is not flat. These drawbacks motivate the search for alternative opto-electronic frequency comb schemes.

## 2.2 Flat-top frequency comb generators

A well-known alternative to generate an optoelectronic comb generator consists of placing in tandem several electro-optic amplitude and/or phase modulators. Figure 2 depicts several possibilities suitable to achieve a comb with a spectrally flat envelope. This type of shape is particularly useful for optical arbitrary waveform generation and optical communication applications (see Sect. 5). Applications requiring different spectral profiles can be obtained by line-by-line pulse shaping a flat-top envelope comb (a technique briefly described in Sect. 3). The challenge in achieving a comb source with flat spectrum lies in setting up correctly the driving characteristics of the modulators. This is not a trivial problem, since even in the simplest case of an electro-optic phase modulator, the system is inherently nonlinear and the power distribution per comb line is not uniform when the RF signal is composed by a single frequency.

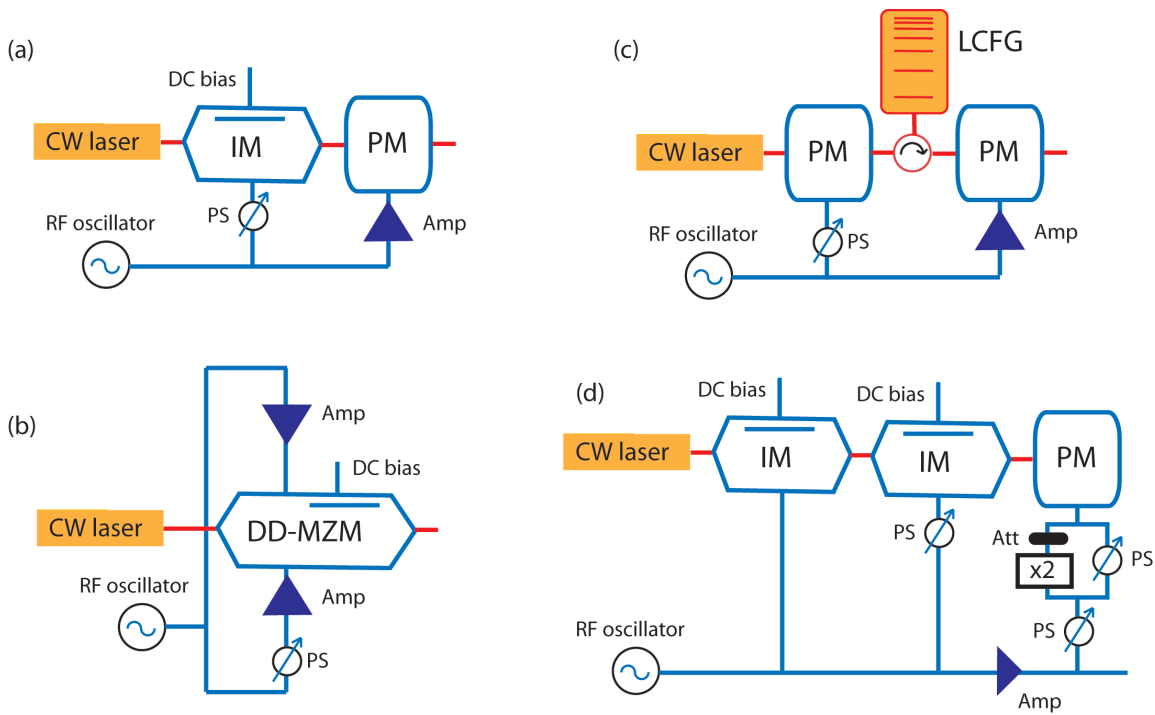

**Fig. 2**. *Layout of common electro-optic frequency comb generators that can provide spectrally flat envelope by an adequate optimization of the settings. IM stands for intensity modulator; PM: phase modulator; DD-MZM: dual-drive Mach-Zehnder modulator; Amp: RF amplifier; PS: electrical phase shifter; LCFG: linearly chirped fiber Bragg grating; Att: attenuator.*

A simple solution to achieve a flat spectrum consists of cascading intensity and phase modulator [30], as illustrated in Fig. 2 (a). The role of the intensity modulator is to produce a train of pseudo-square pulses when biased appropriately. The pulses carve the CW laser only where the chirping provided by the phase modulation is almost linear, thus equalizing the spectrum [13]. Here, the optical bandwidth of the comb is proportional to the modulation index introduced by the phase modulator. Regarding the flatness, this comb can produce a power variation within 5 dB over >50% of the bandwidth (defined at -10 dB). A more recent approach has shown a power variation of 1 dB over ~70 % of the bandwidth by overdriving the intensity modulator [31].

Another alternative to achieve flat-top combs consists of using a dual-drive Mach-Zehnder modulator [32] [33] [34], as sketched in Fig. 2(b). The idea is that with suitable settings for the relative optical and RF phases of upper and lower interferometer arms, the optical spectra obtained from phase modulation are non-flat but become complementary, so that after recombination they provide a relatively flat optical frequency comb. In practical terms this requirement translates into $\Delta\theta_1 - \Delta\theta_2 \pm \Delta\emptyset = \pi$ , where $\Delta\theta_i$ is the modulation index acquired through phase modulation in the corresponding arm and $\Delta\emptyset$ the optical phase difference. Essentially the flatness is the same as the scheme in (a). However, we note that the scheme in (b) requires twice the RF power to achieve the same optical bandwidth.

A more recent approach to achieve optical frequency comb generators with a user-defined envelope is by using the notions of time-to-frequency mapping [35]. The idea can be considered as a generalization of the concept presented in Fig. 2(a) and is implemented in a two-step process, as illustrated in Fig. 3. First, the target spectral profile is tailored on the temporal envelope of a pulse train. In a second stage, another device introduces pure quadratic phase modulation on each pulse (this device is also known as a time lens [36]). If the amount of chirp is sufficiently large, the optical spectral envelope will be a scaled replica of the tailored intensity pulse [37]. This process translates the complexity of the comb synthesis to the design of a pulse train with the desired temporal characteristics and the availability of a repetitive time lens.

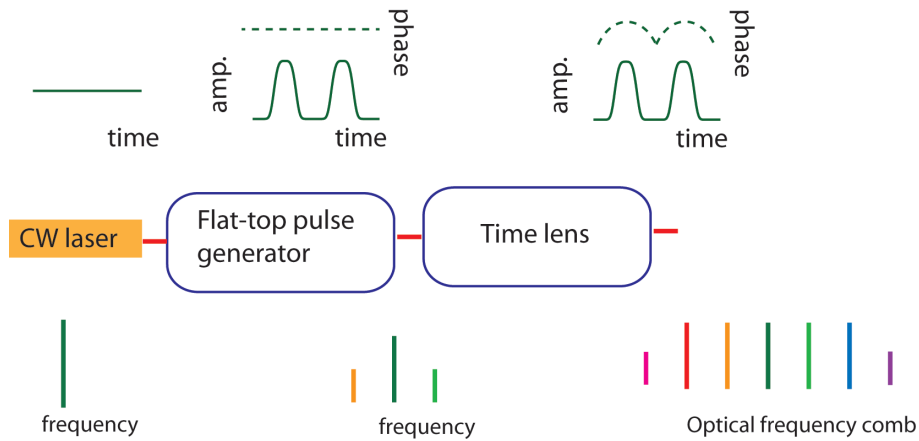

**Fig. 3**. *General layout for flat-top optical frequency comb generation based on time-to-frequency mapping.*

An example of the above scheme can be found in [38], and is illustrated in Fig. 2(c). The first phase modulator and the linearly chirped fiber Bragg grating are configured to provide a pseudo flat-top pulse train with nearly 50 % duty cycle [39]. The subsequent phase modulator is just driven by a single-tone signal. As in the case of Fig. 2(a), the phase modulator introduces an almost linear chirp within the duration of the pulse. Because the pulse train achieved after phase modulation and dispersion with the LCFG is exactly the same as the pulse train achieved with the intensity modulator in Fig. 2(a) [35], these two schemes produce identical types of optical frequency combs. The advantage of this configuration is that it works with just phase modulators, which do not require a DC bias. Thus this platform does not require an active stabilization feedback to compensate for the long-term bias drift. An example from this configuration is presented in Fig. 4 (left). This comb shows a high-quality spectrum with very good flat-top envelope profile [38].

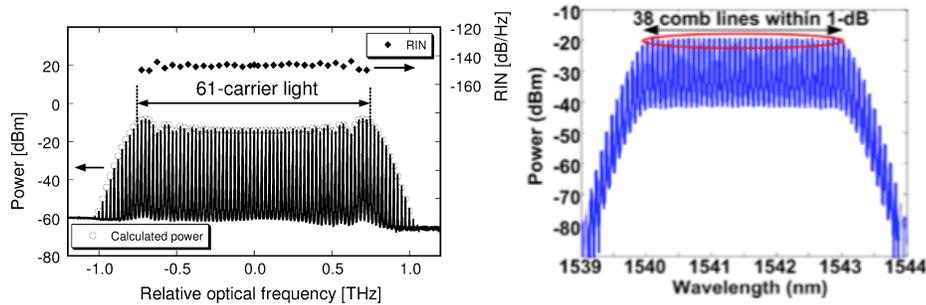

**Fig. 4**. *Experimental examples of flat-top frequency comb generators. (left) corresponds to the approach described in Fig. 2(c). Here, fr=25 GHz and two PMs are used, leading to an accumulated modulation index of 9.7 π. The comb provides more than 60 carriers within 7.4 dB variation. (right) corresponds to an example of comb configured using the approach described in Fig. 2 (d). It is designed to operate at 10 GHz repetition rate and also makes use of two PMs, but the total modulation index in this case is 6.6 π, which explains why there are less comb lines. However, in this case, the flatness is significantly improved thanks to the time-to-frequency mapping approach. Figure on the right reprinted with permission from the Optical Society of America. Results adapted from [38] and [41], respectively.*

Nevertheless, understanding the above comb generation process with the notions of time-to-frequency allows us to assess the limitations in the achievable flatness and provide solutions to improve it. We note that the "rabbit-ear"-like structure at the outer part of the spectrum from Fig. 4(left) is due to the deviation of the phase modulation from a pure quadratic signal within the pulse duration available from the flat-top intensity pulse [35]. The route towards flatter combs becomes clear now. One can either linearize the chirp by shaping the RF signal that is driving the phase modulator, as discussed in [40], or reduce the temporal duration of the flat-top pulses [41]. The idea behind the scheme in Fig. 2(d) exploits both concepts simultaneously [41]. Each intensity modulator is driven to produce a pseudo flat-top pulse, as in Fig. 2(a). When inserted in cascade, the pulse duration is reduced. Importantly, this approach is highly scalable, i.e., larger modulation indexes will broaden the spectrum without altering the overall shape [35]. As a result, a significant increase in the flatness is achieved, as shown in Fig. 4(right). This optical frequency comb has a 1-dB power variation within 83 % of the bandwidth (at -10dB).

Other electro-optic modulation schemes have been reported with a relative good flatness, including driving phase modulators with several RF frequencies [42]; or optimizing the bias and phase shift parameters for a particular RF frequency using two IMs [43]. However, the achievable bandwidth is not as easily tunable with the driving RF power as in the time-to-frequency-mapping method.

## 2.3 From CW to femtosecond-pulse generation

Many applications need broader combs. An advantage of the previous comb generators is that in the time domain, they provide a source of high-repetition-rate optical pulses in the ~1-10 ps regime and can thus be used to achieve nonlinear broadening in a highly nonlinear passive medium. For this, we have to pay some attention to the spectral phase distribution of the comb spectra. In all of the schemes reported in Fig. 2, the phase is close to a parabola and it starts to deviate from this ideal quadratic behavior at the extremes of the spectrum. This means that a passive dispersive medium can be used to compensate for most of this spectral phase and get the desired source of picosecond optical pulses.

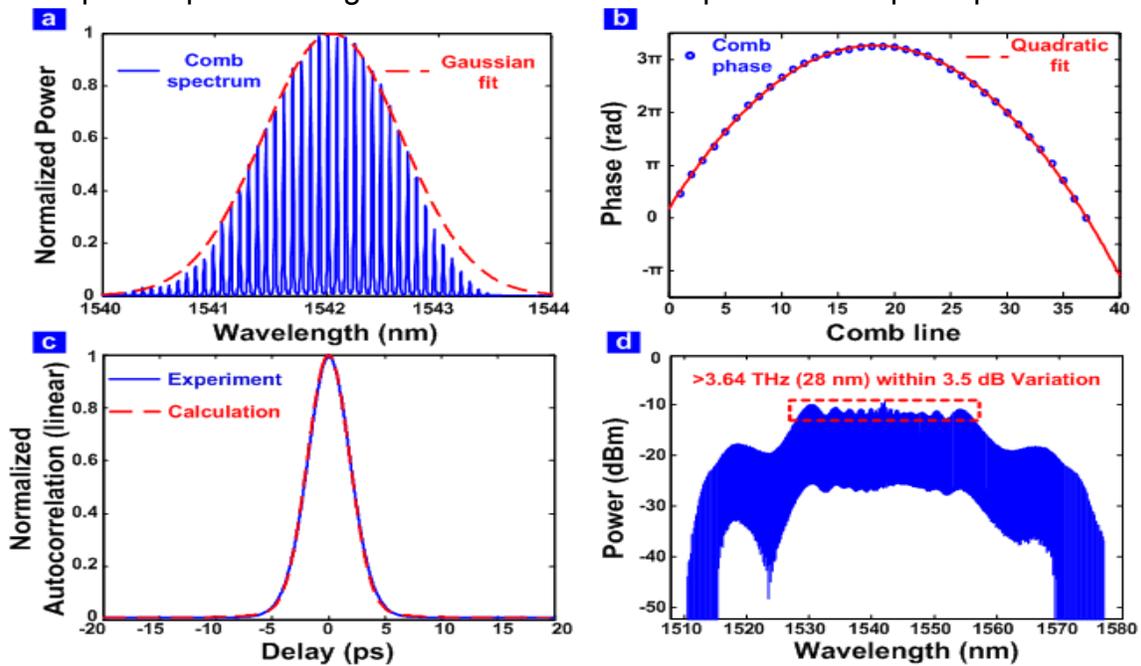

**Fig. 5**. *Spectral broadening of opto-electronic frequency comb generators. (a) Gaussian frequency-comb spectrum (linear scale) generated using the notions of time-to-frequency mapping, as explained in [49]. The spectral phase of this comb is very close to quadratic (b) and it can be easily compensated for using a piece of single-mode fiber. The compressed ultra-short pulse whose autocorrelation trace is displayed in (c) can then be amplified and sent to a highly non-linear fiber with normal dispersion to produce significant coherent broadening and a flat spectral envelope (d). Reprinted with permission from the Optical Society of America. Results reproduced from [48].*

However, maintaining a high level of spectral coherence and uniformity across the spectrum by nonlinear broadening of picosecond pulses is challenging. The underlying reason is that modulation instability may amplify minute amounts of input noise, leading to a degradation of the spectral coherence of the comb [44]. For this reason, the dispersion properties of the external non-linear fiber must be carefully designed. Typically, non-linear fibers with a dispersion-decreasing comb-like profile provide significant broadening (covering the communications C-band) [45] [46] and pulse shortening through adiabatic soliton compression.

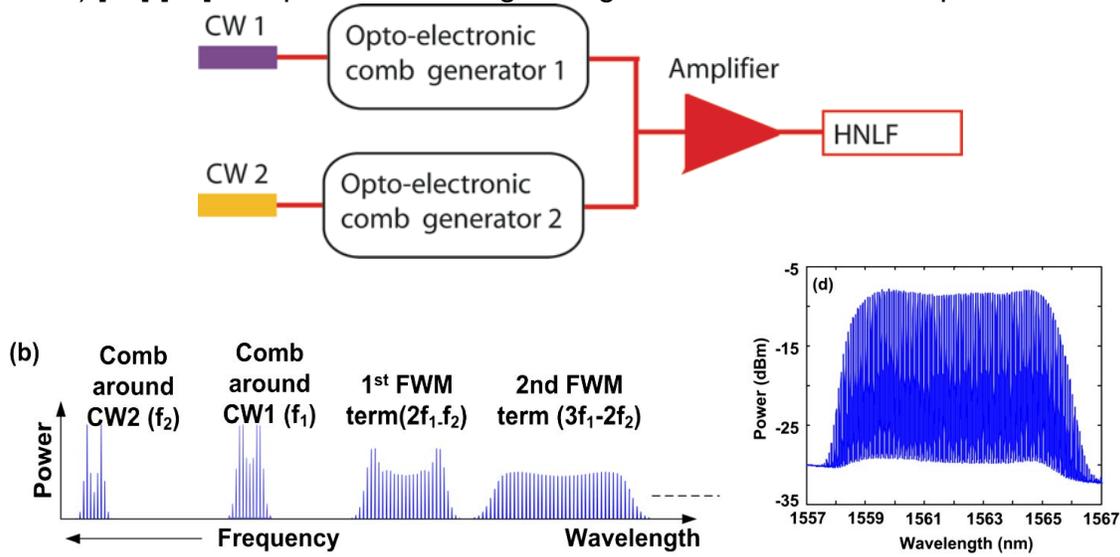

**Fig. 6**. *Spectral broadening of opto-electronic frequency comb generators using four-wave mixing. (a) Schematic representation. (b) When going to lower frequency sidebands, the new generated frequency combs are broader and flatter than the initial seed combs centered at CW1 and CW2. (c) Example of a flat and broad frequency comb generated by exploiting the $2^{nd}$-order sideband and using as the seeds a simple electro-optic phase modulator (around CW2) and a phase and intensity modulator around CW1. More than 100 lines with 10-GHz spacing out of which 75 are within 1dB bandwidth are successfully generated. Figures reproduced from [55] with permission from the Optical Society of America.*

An alternative to achieve broadening is by using highly nonlinear fibers with normal dispersion profile. In this case, modulation instability is avoided and the supercontinuum spectrum shows low-noise performance when the amount of dispersion is sufficiently high [47]. Compared to the adiabatic soliton compression approach, the dispersion slope does not need to be carefully optimized in order to obtain a coherent spectrum, although this comes at the expense of a reduction in broadening factor [44]. In addition, the optical spectrum can be very smooth if the input seed pulse is tailored to have a nearly Gaussian profile [48]. A flat optical frequency comb covering the entire C-band can be achieved by designing an opto-electronic frequency comb generator with a Gaussian profile using the notions of time-to-frequency mapping [49]. Figure 5 illustrates these recent achievements. When the pulse propagates in the normal dispersion regime it may lead to optical-wave breaking [50], whereby different frequency components generated by self-phase modulation meet and interfere at the same temporal location. The occurrence of optical-wave breaking is accompanied by an enhancement of the flatness in the central region of the spectrum [51]. This can be understood because the pulse at the center reshapes into a square-like waveform while the chirp is linearized. The generated spectrum can be compressed to the transform-limited duration only if one manages to compensate for the quadratic spectral phase in the central region and the nonlinear phase of the wings. This task can be accomplished with the aid of a programmable pulse shaper [47] [52]. In a more recent work, the Gaussian pulse necessary to achieve the wave breaking effect was obtained from an electro-optic comb generator followed by cascading a couple of nonlinear optical loop mirrors [53] prior to pumping the non-linear fiber [54]. More than 1500 lines at 10 GHz repetition rate were achieved with a very flat spectrum.

Finally, even flatter combs can be obtained by exploiting a nonlinear effect called four-wave mixing (FWM). Here, the idea is to leverage the $\chi^{(3)}$ response of a nonlinear medium and reshape the complex field of an input waveform in order to achieve the desired characteristics in the spectral envelope. The concept is illustrated in Fig. 6. Two optical frequency combs that are centered at different wavelengths are mixed in an HNLF. Individually, each initial comb has narrow bandwidth and poor flatness. The nonlinear interaction inside the HNLF produces new frequency combs centered at other frequency components through a cascade of FWM. When properly phase-matched, the $N$th-order lower frequency sideband has a complex envelope proportional to $[e_1(t)]^{N+1}[e_2(t)^*]^N$ , where $e_1(t)$ and $e_2(t)$ represent the complex field envelopes of the seed combs. If the seed combs are properly designed, this power enhancement can be exploited to achieve spectral broadening and flattening simultaneously [55]. In the example of Fig. 6, the first comb is formed by cascading an intensity and phase modulator, whereas the second one is a simple phase modulator. This system can provide an effective increase of the modulation index by a factor of $2N + 1$. In the example of Fig. 6, $N = 2$, and the sideband produces a comb with a bandwidth equivalent to what would be obtained with 5 phase modulators. Although certainly promising, we note that this technique would require the seed combs to be frequency-locked in order to get a broader comb retaining long-term frequency stability for the individual comb teeth.

Along these lines, it is worth noting the unique feature of opto-electronic frequency comb generators regarding the possibility to synchronize to (or from) other RF sources. In a recent experiment [56], an electro-optic comb generator with 25 GHz repetition rate was generated using the method displayed in Fig. 2 (a) to achieve >20 nm bandwidth, likely by using a cascade of phase modulators. The clock from the comb was used to synchronize with an optical gate and reduce the repetition rate to 250 MHz. After optical amplification, and compression, the peak power of the pulse was sufficiently high to pump a photonic crystal fiber and produce an octave-broadening spectrum. The generation of an octave-broadening spectrum starting from a CW narrow-linewidth laser is certainly a remarkable achievement. Indeed, provided that the growth of noise can be kept under control, it opens a route to achieve CW self-referencing using the f-2f configuration successfully developed for mode-locked lasers [4]. Rather than for RF photonics, this would be beneficial for frequency metrology applications. However, to our knowledge, the measurement of the beat note corresponding to the carrier-envelope-offset frequency in an opto-electronic frequency comb generator using this approach has not been reported yet.

In a different type of application, the possibility to synchronize an opto-electronic frequency comb generator with an external RF source has been exploited recently for coherent Raman scattering (CRS) microscopy [57]. CRS is a very attractive microscopy modality but it requires two (ideally) picosecond pulse sources centered at different wavelengths (one of them tunable) with synchronized repetition rates. In [57], an opto-electronic frequency comb at 1 $\mu$m was generated by driving with the RF signal derived by photodetecting a low-repetition-rate Ti:Sa laser. This example is a clear illustration of the potential of opto-electronic frequency comb generators in terms of the flexibility in tuning the CW laser and the repetition rate of the source.

## 2.4 Towards higher repetition rates with parametric frequency comb generators

We have seen that opto-electronic frequency combs generated by electro-optic modulators are very flexible in their setting characteristics but their repetition rate is limited by the bandwidth of the modulator. As mentioned before, with commercially available modulators made of lithium niobate, this imposes a limit of ~ 40 GHz. However, higher repetition rates can be achieved while still exploiting the characteristics of electro-optic frequency comb generators. The idea is drawn in Fig. 7. Two lines from an initial frequency comb generator are selected by an arrayed waveguide grating and used to injection lock two remote different lasers centered at different wavelengths. Because the comb maintains the phase coherence across the whole bandwidth, the lasers will inherit the phase stability from the parent lines. The advantage of this technique is that the spectral gap between the two lasers can be set with the same accuracy as the RF oscillator. These two phase-locked frequencies are subsequently amplified and sent to a highly nonlinear fiber to induce a cascade of new frequencies that are generated by four-wave mixing. In this way, a frequency comb emerges with the repetition rate given by the difference in frequency between the two parent lasers. Thus, the ultimate limit in repetition rate is not fixed by the RF oscillator anymore but by the optical bandwidth of the initial seed comb [27].

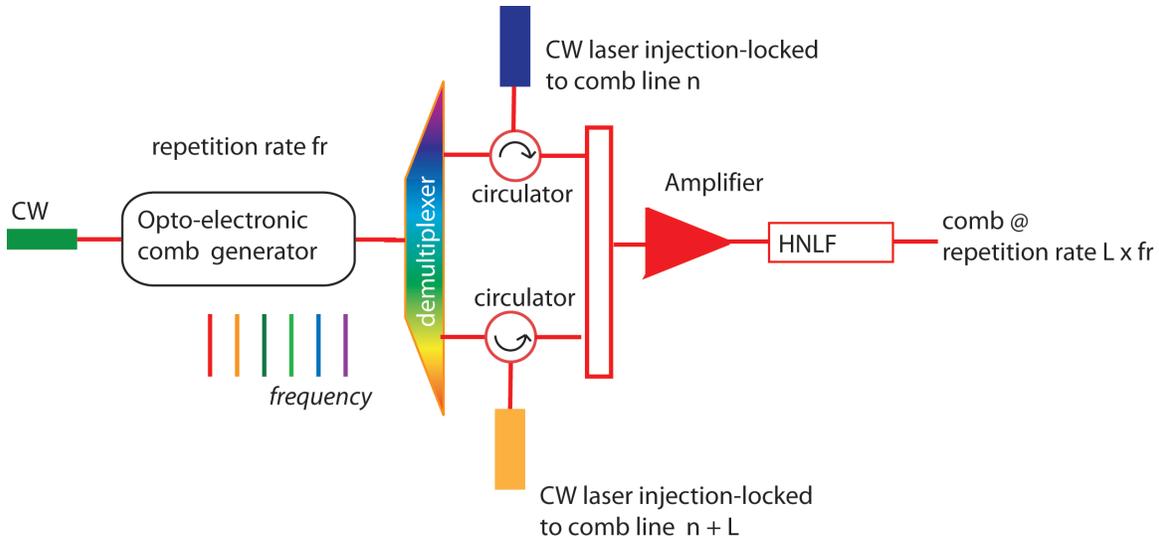

**Fig. 7**. *Scheme to achieve ultra-high repetition rates using injection locking and FWM effects in nonlinear media.*

With respect comb-line filtering alone, the use of injection-locked seed lasers is more advantageous because it can provide higher OSNR and input power to drive the required amplifier. It is also important to note that in order to achieve a significant number of lines, phase matching across a broad bandwidth must be obtained but more importantly, the HNLF should avoid the effect of stimulated Brillouin scattering (SBS) [58]. The SBS power threshold can be increased by e.g., applying strain or a temperature gradient on the fiber or by doping the core (e.g. with aluminum). However these processes usually modify the dispersion characteristics of the fiber, which in turn affects the phase matching. Nevertheless, recent results have shown frequency combs at >100 GHz repetition rates [27], as highlighted in Fig. 8. Other groups have shown that further broadening can be achieved by cascading two FWM stages [59]. Interestingly, here only the first HNLF needs to have a high SBS threshold and the second one is optimized to achieve a significant amount of broadening.

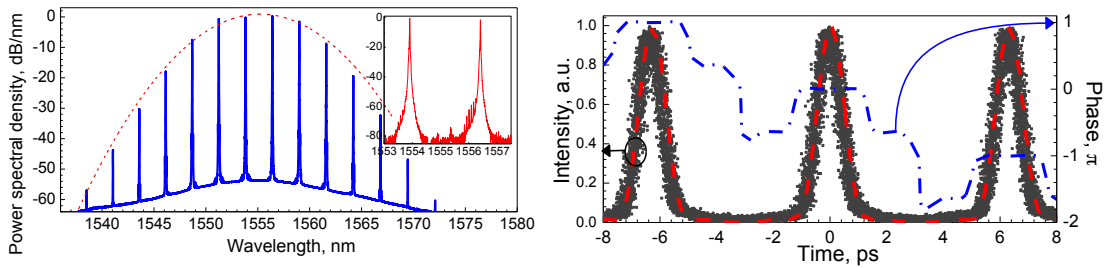

**Fig. 8**. *Repetition-rate-reconfigurable frequency comb using the technique presented in Fig. 7. Several lines spaced by 320 GHz are achieved with very high contrast and OSNR (left). The phase stability of this comb generator is demonstrated by showing temporal traces of high-speed pulses (at 160 GHz rate) measured with an optical sampling oscilloscope (right) with sub-ps temporal resolution. Results are adapted from [27]. Figures courtesy of R. Slavik.*

# 3. Brief overview on Fourier processing of optical frequency combs

Fourier pulse shaping is now a ubiquitous technique in fields as diverse as optical communications, nonlinear microscopy, coherent control and quantum information processing [60]. This technique realizes ultrafast waveforms with amplitude, phase, and/or polarization characteristics defined by the user. It works by the manipulation in parallel of the frequency components of an input broadband waveform and leverages the flexibility of spatial light modulation technologies to achieve user programmability. Since its inception, this technology has evolved to the point that the spectral resolution may be comparable to or better than the repetition rate of the input pulse train [61] so that its discrete frequency components can be manipulated individually. This fundamentally new regime, in which the input waveform is processed in a "line-by-line" manner, opens up unique possibilities by merging the flexibility of Fourier pulse shaping with the superb phase-noise performance of broadband frequency comb sources [9].

The reader looking for introductory material into the general field of pulse shaping can find comprehensive review articles written by one of the present authors in [60] and [62]. The work in [63] reviews the fundamentals of line-by-line pulse shaping and applications in optical and RF arbitrary waveform generation prior to 2008. In this section, we shall briefly review the most recent technical efforts realized towards the Fourier processing of frequency combs.

In the optical communications C-band, line-by-line pulse shaping may be achieved by combining fiber optic and free-space components [61]. A recent approach [64] uses a programmable two-dimensional (2D) liquid crystal on silicon (LCOS) display in a reflective configuration. The LCOS display allows for gray-level amplitude and phase control and the setup achieves ~ 10 GHz spectral resolution and a bandwidth spanning the whole C-band. The device is rather compact and shows ~4-5 dB loss when no spatial mask is programmed. This resolution level allows for line-by-line shaping of high-repetition-rate optical frequency combs. A key novelty of this arrangement is the use of a cylindrical lens to focus the different spectral components. The lens replicates the input spectrum on to several rows of the LCOS display. Then, the user can program a vertical grating on the display to steer different portions of the spectrum to selected output fiber ports. This provides a spatial extra-degree of freedom in pulse processing [65].

The complexity of a waveform synthesized by any pulse shaping apparatus can be quantified by means of the time-bandwidth product (TBWP). This figure of merit is defined by the ratio between the bandwidth of the shaped pulse to the shaper resolution. In line-by-line pulse shaping, the TBWP provides an estimation of the number of comb lines that can be processed individually and is ultimately limited by the number of addressable elements in the shaper. Waveforms synthesized in a line-by-line manner with 5 GHz resolution have shown TBWPs exceeding 100 [66]. To increase further the TBWP, researchers have combined a prism with a grating to introduce the angular dispersion required to spread the frequency components in one dimension. The prism rectifies the nonlinear dispersion of the grating so that the modes appear evenly spaced in the Fourier plane across the whole bandwidth. This in turn optimizes the use of the pixels available in the 1D display. With this grism-based pulse shaper, researchers have shown control over 600 comb lines at 21 GHz resolution [67]. In contrast, two-dimensional pulse shaping [68] is a new regime where the frequency components in the Fourier plane are spread over the two dimensions (as opposed to only one) by crossing two angular dispersive elements, i.e., a virtually imaged phase array (VIPA) [69] and a grating [70]. This pulse shaping modality permits to increase the TBWP of the synthesized waveform by several orders of magnitude by leveraging the large number of available pixels in 2D displays [68]. Recent demonstrations have shown a programmable 2D pulse shaper with more than 1000 addressable elements [71] and ~3 GHz resolution.

Regarding integration, pulse shapers with a resolution in the order of tens of GHz can be achieved by using arrayed waveguide gratings as spectral dispersers/combiners inscribed in a planar lightwave circuit made of silica [72] [73] [74]. The different frequency components can be manipulated in amplitude and phase exploiting thermo-optic modulation in the waveguide. Recent efforts consider integration in semiconductor structures. Here, static shaping of 100 frequency comb components at 10 GHz repetition rate has been recently reported using quantum-well phase modulators monolithically integrated in an InP platform [75]. Two modulators per channel are designed in order to achieve independent amplitude and phase modulation.

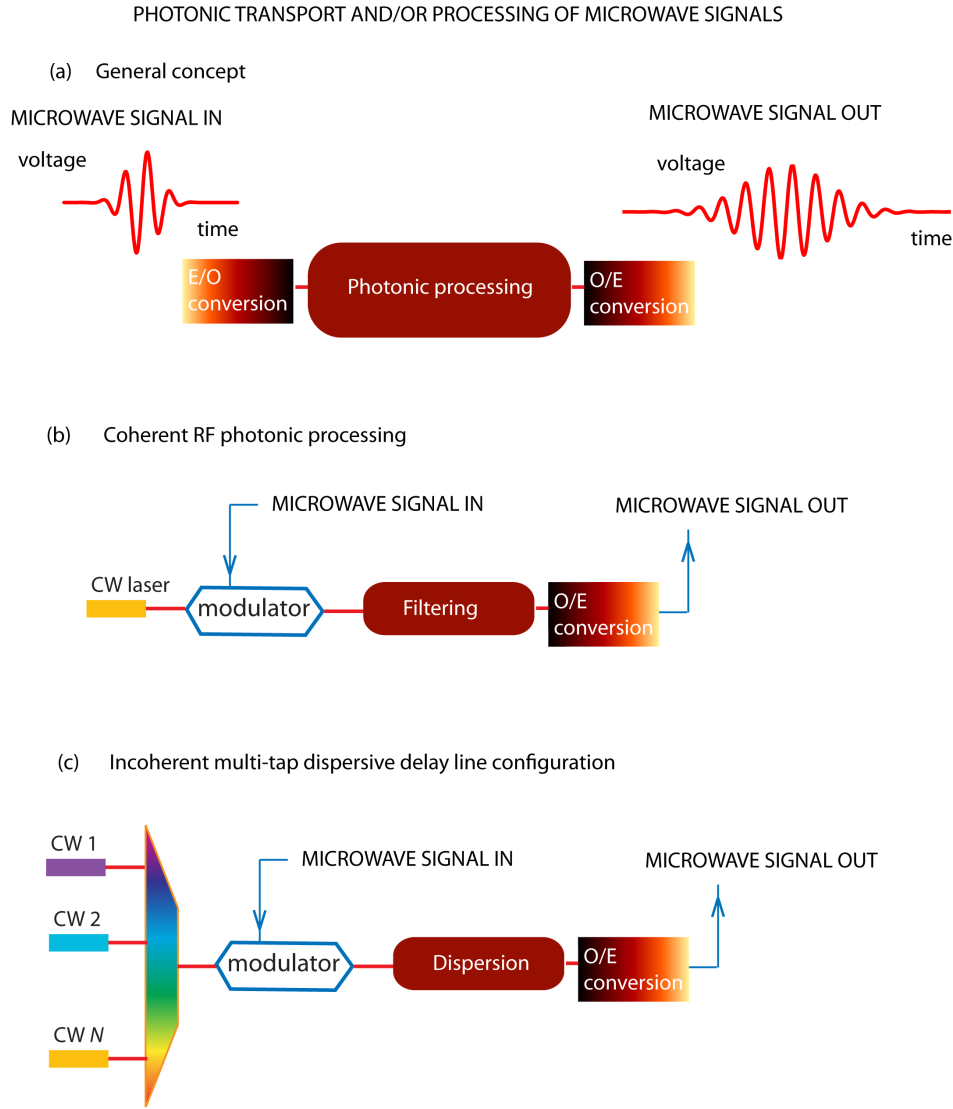

**Fig. 9**. *Photonic transport and/or processing of RF signals: (a) General scheme. An E/O converter maps the input RF signal to the optical domain, where is manipulated to be finally brought back to the RF domain via high-speed photodetection. In the case of an RF filter, this process can be implemented in a coherent scheme (b) or an incoherent multi-tap dispersive delay line configuration (c). The latter approach can benefit from an opto-electronic frequency comb generator.*

# 4. RF signal processing applications

The advent of low-loss single-mode fibers made it possible to transmit RF signals over much longer distances than what is possible with microwave waveguides. The transmission of RF signals (either analog or digital) over fiber plays a significant role in applications like wireless communications, military radar or radio astronomy. These applications demand highly linear links with large dynamic range and signal-to-noise ratios, and are at the core of the field of RF photonics. Reference [2] reviews the evolution of microwave photonic links and related devices up to 2006. The device parameters affecting the link performance and the figures of merit used to quantify it are described in a comprehensive way in reference [76].

In essence, a microwave photonic link is composed by three parts, as sketched in Fig. 9(a). The content of the input RF signal is transferred into the sidebands of an optical carrier by means of some electrical-to-optical (E/O) conversion process. The optical signal is then sent through fiber and transferred back to the electrical domain with a high-speed photodiode. Ideally, the process E/O conversion → optical processing → O/E conversion must be done in an efficient way, preserve the linearity of the link, and minimize the amount of noise introduced through the process. This introduces formidable technical challenges for the devices involved [77]. Nevertheless, recent efforts have shown microwave links with gain without using any RF amplifiers and having a noise figure <10 dB over several GHz of bandwidth. This achievement was possible thanks to a wise architecture employing balanced detection to reject common noise sources [78].

In the early years of microwave photonic links, it was soon recognized that photonics technologies could provide an added value to the microwave link by performing signal processing tasks directly in the optical domain [79] [80]. The motivation to do so lies in the fact that photonic technologies can handle significantly higher bandwidth and provide a level of flexibility and programmability unattained by microwave counterparts. Today, we understand that the synergy between state-of-the-art components in microwave engineering and photonic technologies offers a truly unique potential to realize RF signal processing tasks that are difficult or just impossible to achieve using microwave technologies alone [1]. Among these, one of the most researched applications is the realization of highly programmable RF filters that operate at high frequency carriers over bandwidths spanning several GHz.

In the simplest configuration of a programmable RF-photonic filter, one can think of an optical filter placed between the E/O and O/E conversion process in the general scheme depicted in Fig. 9. The optical filter acts on the optical carrier and one of the sidebands generated upon E/O conversion. In the literature, this configuration is often termed as *coherent filter* [81]. The most important attribute of this setup is that the RF filter transfer function becomes a replica of the optical filter. The challenge of this coherent approach is the need for a high-quality optical filter with ultra-fine resolution and user programmability. Recently, a VIPA-based pulse shaper with sub-GHz resolution was used as a reconfigurable optical filter to provide an RF photonic filter with user-defined amplitude and phase characteristics [82] [83].

## 4.1 Multi-tap dispersive delay-line RF photonic filters

A different architecture for microwave signal processing is the one based on a tapped delay-line configuration [84] [85]. Mathematically, the filtering operation is defined by a finite impulse response (FIR) with programmable tap coefficients and delays. The most flexible arrangement corresponds to a dispersive delay line configuration [86]. Here, a multi-wavelength source is externally modulated by the RF signal to be filtered, sent through a dispersive medium and detected by a photodiode. The different modulated carriers travel at a different speed in the medium and therefore arrive at different instant times. The wavelength separation between two consecutive carriers must be larger than the photodiode's bandwidth, so that the current generated is composed of a summation in intensity of the different contributions from each of the optical carriers [85]. The main difference of this platform with respect to the coherent approach is that the filter characteristics can be easily controlled with great accuracy through the physical parameters of the optical source and the dispersive medium. More concretely, the RF impulse response can be written as

$$h(t) \propto \sum a_n \delta(t - nT), \tag{1}$$

where the coefficient $a_n$ is established by the power carried by the $n$th optical carrier and the tap delay $T = \Phi_2 \Delta\omega$, with $\Phi_2$ being the group-delay-dispersion coefficient of the medium (in time square units) and $\Delta\omega$ the angular frequency separation between two consecutive optical carriers. Thus, by controlling the frequency separation and the optical power distribution of the source in a tap-by-tap manner, the filter's response can be synthesized according to Eq. (1). The reader looking for further details on the different photonic architectures, design parameters and limitations of FIR RF photonic filters is referred to the tutorial papers [85] and [87].

For the above architecture, having a multi-wavelength source with phase coherence across the bandwidth is unimportant. In fact, it has been implemented with success using either coherent or incoherent light sources, including an array of laser diodes [86], an apodized broadband incoherent source (see e.g. [88] [89] [90]), or actively mode-locked lasers [91]. However, using an opto-electronic frequency comb generator still provides unique advantages [92] [93]. The most straightforward one being the fact that only one laser is needed to produce multiple optical taps. Using the comb sources presented in Sect. 2.3, microwave photonic filters with hundreds of taps become readily available, providing an unprecedented degree of flexibility when combined with line-by-line pulse shaping [92]. It is interesting to note that, unlike with the coherent scheme, in this approach the spectral resolution of the microwave filter can be much finer than the resolution of the optical filter used to apodize the optical source. Regarding noise performance, because the taps are narrow-linewidth, they do not generate the undesired beat noise terms observed in experiments based on spectral slicing of broadband incoherent light [94]. In a recent experiment [95], a microwave photonic filtered link implemented with an opto-electronic comb source has shown a noise performance close to that achievable with a single narrow-linewidth laser. It is important to note that any MWP filter based on an FIR multi-wavelength dispersive delay-line geometry will have a spur-free operation range given by half the frequency spacing. For tunable multi-wavelength sources (e.g. a bank of laser diodes), this spacing can be quite large, whereas for optoelectronic comb generators the spacing is equal to the repetition rate of the comb. This limits the spur-free RF operating range to ~ 20 GHz for electro-optic comb generators, but with the parametric comb generators presented in Sect. 2.4 the operating range can be set to be > 100 GHz.

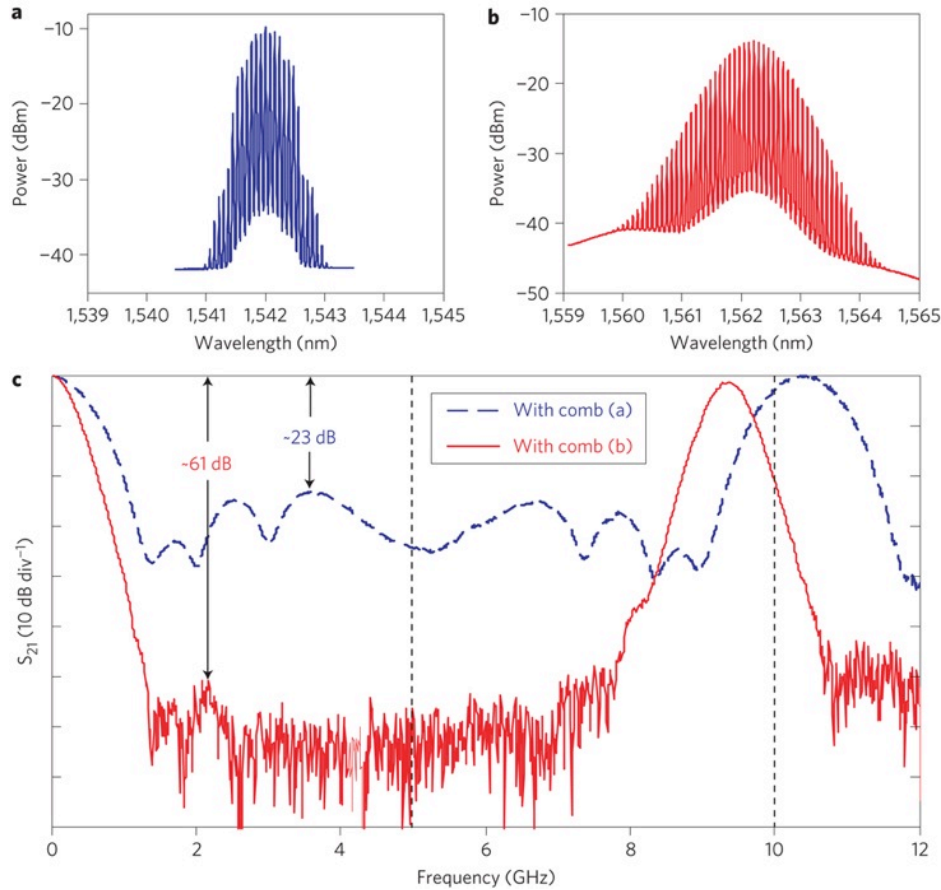

**Fig. 10**. *Ultra-high sidelobe suppression microwave photonic filtering exploiting high-repetition-rate optical frequency combs. The initial comb in a) is broadened and smoothed using the FWM technique described in Sect. 2.3 to achieve a Gaussian-apodized comb like the one presented in b). The tap weight distribution from this comb source is ideal to realize to realize a microwave photonic filter using a multi-tap dispersive delay line configuration. Results reproduced from [93] with permission from McMillan Publishers.*

Using the notions of time-to-frequency mapping highlighted in Sect. 2.2, the desired tap weight distribution can be obtained without the need of a pulse shaper [49]. In addition, the FWM techniques pointed out in Sect. 2.3 can be exploited to smooth the optical spectral envelope and obtain the tap weight coefficients necessary to achieve a filter with a sidelobe suppression >60 dB, as shown in Fig. 10. Except for one notable exception [96], this out-of-band rejection level has been rarely reported with microwave photonic filters. It is worth noting that in order to achieve this level of performance, the tap weight coefficients must remain stable in power to less than 0.2%.

Yet, probably the most interesting attribute of opto-electronic frequency comb generators for RF photonic filters is the fact that their characteristics can be synthesized in a very rapid manner through external electrical control signals [93]. This permits to allocate the filter's central frequency with high precision. In the following subsection, we shall show a very illustrative example of this.

### 4.2 Phase-programmable filters

A clear limitation from a multi-tap dispersive delay-line microwave photonic filter is that, in principle, the weight coefficients in the FIR of the microwave filter can only be synthesized to be positive. Ideally, one would like to be able to program the tap weights from Eq. (1) with complex-valued user-specified values. This level of flexibility would enable the user to tune the filter bandpass even away from the baseband, as well as to program the amplitude and phase response of the filter in a completely independent manner. However, having complex-valued coefficients is challenging. Physically, a scheme to control the relative phase between the optical carriers and the generated sidebands is needed. At input RFs with high-frequency carrier, this can be done by placing a programmable optical filter with sufficient spectral resolution after the modulation stage, as reported in [97] [98]. Other schemes make use of Brillouin scattering effects in fibers [99] or semiconductor chips [100], but then the extension to multiple taps becomes very challenging.

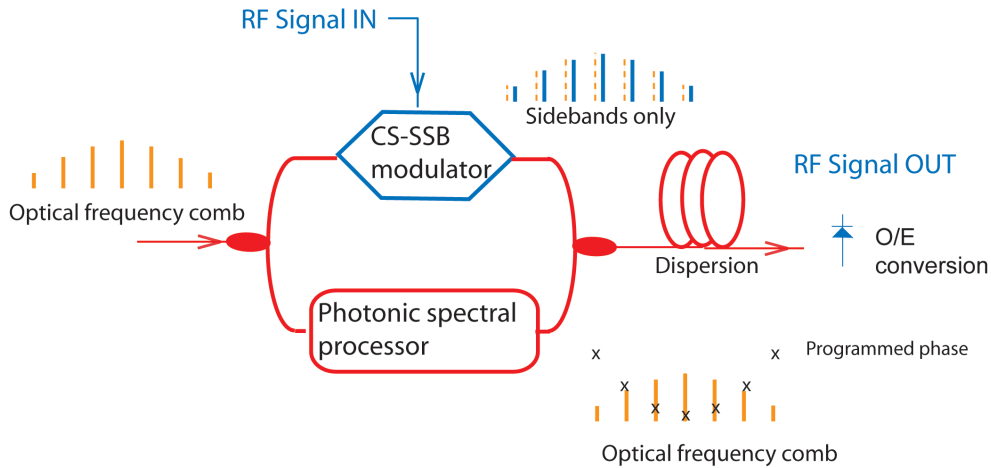

**Fig. 11**. *Schematic representation of a microwave photonic filter with programmable complex-coefficient taps implemented with an optical frequency comb in a dispersive delay-line configuration.*

A relatively simple solution, firstly reported in [92], consists of using an opto-electronic frequency comb generator in an interferometer configuration like the one presented in Fig. 11. Here, the individual optical taps are manipulated in amplitude and/or phase in their pass through the lower arm. In the upper arm, the carriers are suppressed and only the sidebands generated by modulation of the RF signal to be filtered remain. Upon recombination, the relative phase between the parent line and optical sideband can be controlled in a tap-by-tap manner. In mathematical terms, it can be shown that the RF filter has an impulse response as the one given by Eq. (1), with the coefficients $a_n = e_{1n}^* e_{2n}$. Here, $e_{1n}$ and $e_{2n}$ represent the complex field of the corresponding $n$th line of the initial and processed comb, respectively. It becomes clear that by controlling the optical amplitude and phase of the lines, complex $a_n$ coefficients can be synthesized on demand. A key aspect is that the initial phase of the comb source plays no role in the RF's filter response.

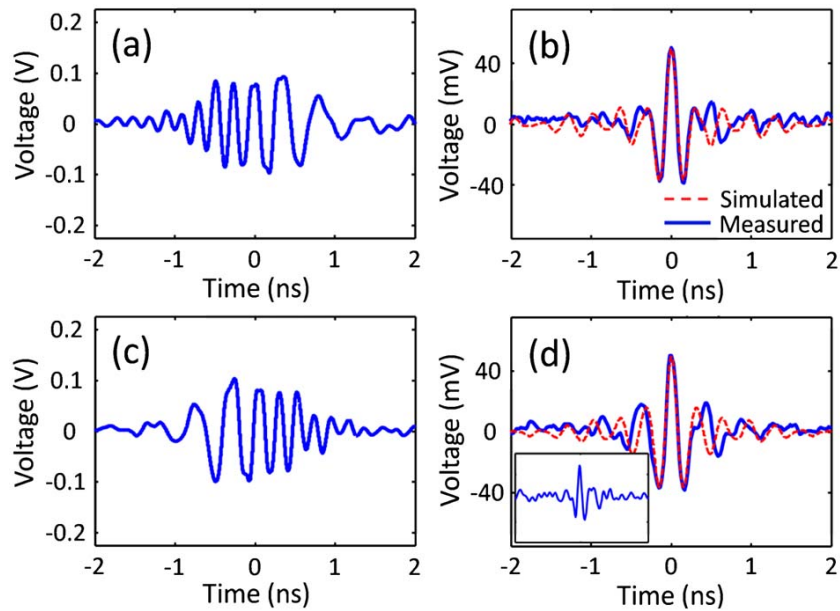

**Fig. 12**. *Reconfigurable phase-matched filtering of an input microwave broadband signal with a microwave photonic filter implemented with an opto-electronic frequency comb generator. A downchirp waveform a) is input to a microwave photonic filter whose phase response is programmed to compensate for that of the input waveform and achieve a transform-limited (zero chirp) waveform as shown in (b). The filter programmability is illustrated in the bottom row results where an input waveform synthesized with the same amount of chirp as in b) but reversed in sign is successfully compressed too. The inset in (d) shows a single-shot waveform with an SNR of ~20dB. Results reproduced from [101] with permission from the Optical Society of America.*

When the photonic circuit in the lower arm is realized by a line-by-line pulse shaper, the microwave filter inherits its programmable characteristics [101]. In [101], we showed an example where a ~ 20-tap microwave photonic filter is synthesized to have a quadratic phase response. This in turn allows for correcting the spectral phase of an input microwave signal. The example in Fig. 12 shows a phase-filtering experiment in which a chirped input microwave signal with ~3 GHz instantaneous bandwidth is synthesized by a state-of-the-art arbitrary waveform generator. Thanks to the programmability of the microwave photonic filter, the input signal can be easily compressed to the transform-limited duration regardless of the sign of the input chirp.

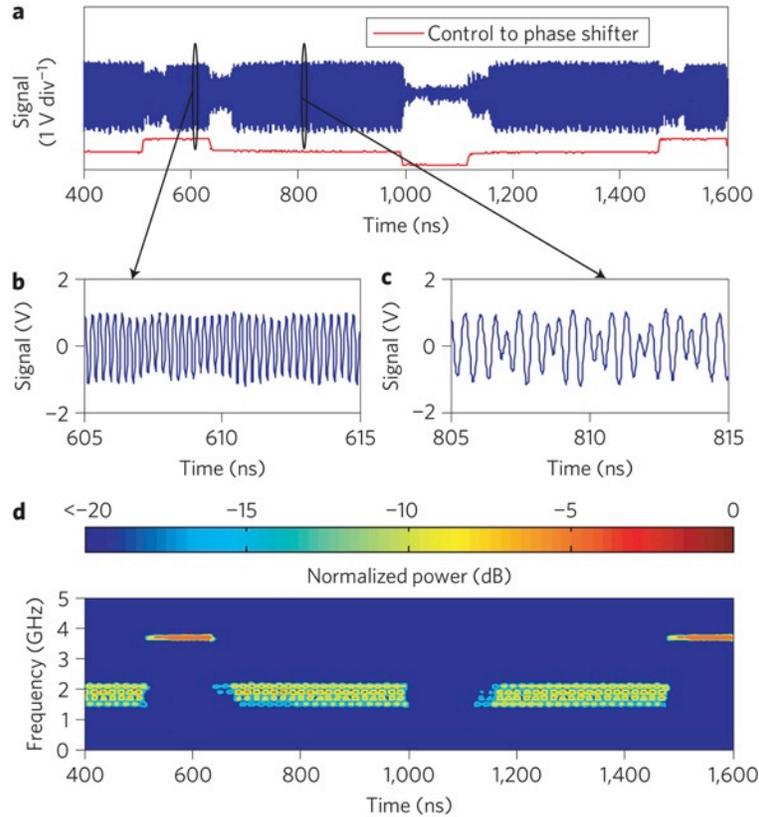

**Fig. 13**. *Dynamic tuning of comb-based RF photonic filters. Two input RF signals, i.e., a CW signal at 3.7 GHz and a 1.8GHz signal carrying BPSK data at 600Mb/s are launched simultaneously into an RF photonic filter. The filter can be tuned at either band by controlling the phase shifter on an optical frequency comb generator. Figures a), b) and c) show the temporal trace of the output filtered signal, together with the control voltage sent to the phase shifter. The dynamic response characteristics of the filter are better visualized with the aid of the time-domain representation in d), where the spectrogram of the temporal trace in a) is calculated. The filter dynamically tunes its band within less than 100 ns and suppresses the undesired bands for more than 20dB. Results reproduced from* [93] *with permission from McMillan Publishers.*

Tuning the filter's bandpass can be easily achieved by programming a linear spectral phase across the optical spectrum in the lower arm [92]. This is equivalent to controlling the relative time delay between the path lengths. It is very interesting to note that with opto-electronic frequency comb technology, this task can be accomplished by controlling the phase shift of the RF signal that drives the comb [93]. The key point of this strategy is that the tuning mechanism is free of mechanical artifacts and thus can be reconfigured at sub-microsecond speed. This is significantly faster than what is possible with most of today's electrically tunable microwave filters. The characteristics of these dynamic filters cannot be assessed with common vector network analyzers and rapid time-domain instruments are required instead. In the experimental results presented in Fig. 13, the bandpass from a microwave photonic filter is tuned in less than 100 ns. The figure presents the filtered output signal and the computed spectrogram. This time-frequency representation helps to visualize the dynamic changes in the filter response.

## 4.3 Photonic downsampling and reconfigurable filtering

An alternative scheme for processing RF signals consists of translating their frequency content to a region accessible by high-speed analog-to-digital converters (ADCs). Then additional digital signal processing can be performed either offline or in real-time. Photonic technologies can help to implement this RF mixing operation at bandwidths unattainable by microwave counterpart technologies.

Probably the most heavily researched scheme involves photonic sampling (also known as subsampling or downsampling) [102], which can be seen as a particular implementation of a photonic ADC [103]. Here, the RF signal to be downconverted modulates the peak amplitude of a train of ultrashort pulses. The RF carrier is expected to be higher than half the pulse train's repetition rate, fr, but its instantaneous bandwidth is restricted to less than fr/2. Then, a photodiode with cut-off frequency fr/2 deliberately introduces aliasing and replicates the frequency content of the RF signal at baseband. In an RF photonic downsampling link the sampling process is realized in real time, i.e., the input RF signal does not need to be repetitive. This means that it can be downconverted in a single shot as long as the above-mentioned fundamental requirements are met. This process can be alternatively understood in the frequency domain, where the optical sidebands generated upon RF modulation beat with the nearest adjacent discrete line from the optical spectrum of the pulse train. It is interesting to note that this RF photonic sampled link has the same performance characteristics as the RF photonic links implemented with a CW laser for the same photodiode current and neglecting photodiode nonlinearities [104]. The link performance in the simplest case of intensity-modulation direct-detection has been comprehensively analyzed in [105].

In photonic downsampling, the sampling is performed in the optical domain and the digitization is realized with electronics. Thus the ADC bandwidth must be equal or higher than the repetition rate of the pulse train. Because the signal is digitized, this type of RF photonic links offers the possibility to improve the linearity by performing additional digital signal processing [102]. Photonic sampling has been explored since the early years of mode-locked lasers [106], but the prospect for ADC lies in the broad bandwidth and low timing and amplitude jitters available from mode-locked lasers [103]. Photonic ADCs with a large effective number of bits (ENOBs) have been reported with passively mode-locked fiber lasers [107], which offer outstanding timing jitter characteristics [108] albeit at the expense of limited repetition rates. In this direction, semiconductor-based harmonically mode-locked lasers offer promising prospects towards higher repetition rates with low timing jitter characteristics [109].

Opto-electronic frequency comb generators can offer interesting advantages for photonic downsampling RF photonic links [110]. The light sources described in Sect. 2 have repetition rates comparable to state-of-the-art mode-locked lasers and are thus capable to downsample RF signals with instantaneous bandwidths approaching ~ 20 GHz. With the frequency comb techniques presented in Sect. 2.5, repetition rates in excess of 100 GHz and ~ 100 fs timing jitter can be obtained [27]. This would translate into downsampling RF waveforms with ~ 50 GHz bandwidth and 7.5 ENOBs if timing jitter was the only limiting factor [103]. In reality the modulator's bandwidth and photodiode nonlinearities would limit performance. The most interesting characteristic of the optoelectronic combs presented in Sect. 2.2 for photonic downsampling is that, unlike mode-locked lasers, their repetition rate is tunable. This means that they can be quickly reconfigured to operate at the RF band of interest with an accuracy provided by the RF oscillator driving the comb. This feature was jointly exploited with the line-by-line pulse shaping techniques of Sect. 3 in order to achieve programmable analog filtering (including bandpass tuning and apodization) prior to the digitization stage in photonic downsampling [110]. The analog filter transfer function in this case is finite impulse response with complex coefficient taps, whose weights depend on the optical spectral phase of the comb.

## 5. High-performance waveform synthesis

We mentioned in Sect. 3 that optical arbitrary waveform generation enables unique applications thanks to the combination of the flexibility available from line-by-line pulse shapers with the extraordinary phase-noise performance across the bandwidth from frequency combs [9]. In this section we will give a flavor of the possibilities that O-AWG can provide in the field of coherent communications (considering both domains optical and wireless).

### 5.1 Synthesis of single-tone high-purity microwave signals

The central concept behind the synthesis of arbitrary microwave and terahertz signals using optical frequency combs is relatively simple. The individual lines from an optical frequency comb are manipulated in amplitude and phase to yield a train of optical pulses with user-defined characteristics. Then, after detection in a high-speed photodiode, the pulse train provides an electric dynamic signal with bandwidth content just limited by the O/E conversion bandwidth [63]. In the simplest case in which the comb is synthesized to yield a train of ultrashort pulses, the spectrum of the microwave signal will be another frequency comb whose ticks are evenly spaced by a quantity matching the repetition rate of the optical pulse train. If the bandwidth of the photodiode is sufficiently large, the frequency grid may enter well into the millimeter-wave or even the terahertz regime when specialty photodiodes are used [111]. In the case of the optoelectronic frequency comb generators described in Sect. 2, the spectral purity of a single line from the photodected microwave comb will be, in the best case scenario, as good as the purity of the RF oscillator used to generate the comb [112]. However, when using carrier-envelope-offset stabilized frequency combs, the quality of the photonically generated microwaves may be better than those available from state-of-the-art microwave oscillators working at room temperature [113].

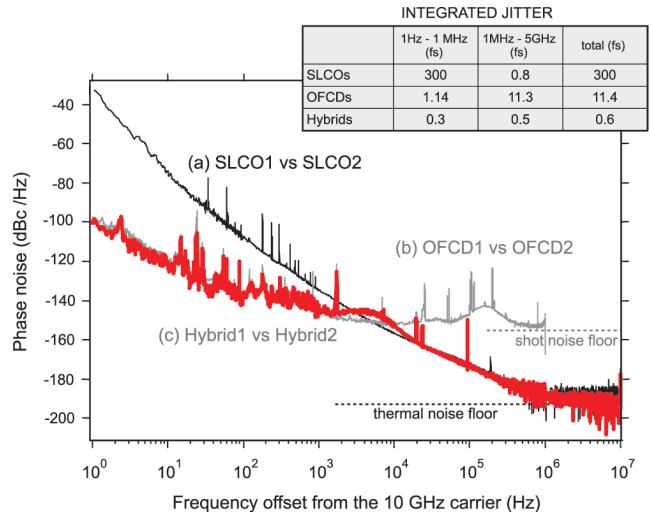

**Fig. 14**. *SSB RF spectrum of the 10-GHz microwave signal generated from: a state-of-the-art SLCO (a); photodetecting and filtering the 10th harmonic of a 1-GHz frequency comb with carrier-envelope-offset frequency stabilization (b); and phase-locking the SLCO to the 10-GHz generated from the comb (c, red curve). The inset Table indicates the integrated phase noise and corresponding timing jitter contributions from each band. Results reproduced from [117] with permission from the American Institute of Physics.*

Achieving these microwave signals is not as easy as it sounds. The purity of a microwave tone is assessed by its single-sideband (SSB) RF spectrum. An ideal sine wave should have a linewidth comparable to the inverse of the observation time period, but of course realistic and sometimes unavoidable noise contributions will broaden further the spectrum. In order to achieve an ultra-pure single-tone microwave signal, first an optical frequency comb is phase-locked to a cavity-stabilized reference CW laser [113]. In this way, the comb inherits the fractional frequency stability of the optical reference, which can approach 1 part in $10^{16}$ in 1-10 s [114]. In a recent work, the microwave signal corresponding to the RF filtering of the 10th harmonic of a photodetected 1-GHz mode-locked Ti:Sa laser with stabilization of the carrier-envelope-offset freuency showed an unprecedented integrated phase-noise performance that beat the purity of the best 10-GHz microwave oscillator operating at room temperature, which is based on a sapphire-loaded cavity (SLCO) [115]. Nevertheless, the improvement appears only at the low-frequency region of the SSB RF spectrum. This is due to the fact that at higher frequencies, the noise performance of photodetected pulse trains is limited by the photodiode's shot noise. Interestingly, in a paper carefully revisiting the origin of shot noise in ultrashort pulse trains, it was recently shown that with specially designed linear high-power-handling photodiodes, the noise floor may be equal or better than the thermal noise floor in SLCOs [116]. For regular photodiodes however, the shot noise is higher than the thermal noise floor. In a recent work, researchers at NIST bypassed the shot-noise-limit constraint by phase-locking the SLCO to the 10-GHz tone generated by the photodetection of a carrier-envelope-offset stabilized comb [117]. Then, for short temporal periods, the resulting microwave signal from the oscillator inherits the performance of the SLCO, whereas for longer operation times it follows the comb-based signal's. As Fig. 14 indicates, the integrated SSB RF spectrum provides an impressive sub-fs timing jitter.

This timing accuracy should enable new applications in high-resolution radar and high-speed sampling. Note that although sub-fs timing jitters have been reported in low-repetition-rate (MHz class) passively mode-locked lasers [108], this is much more difficult to achieve at higher repetition rates because strictly, the SSB RF spectrum of the microwave signal must be integrated over more decades to reach the Nyquist limit.

**5.2 Dynamic radio-frequency arbitrary waveform generation**

Arbitrary RF waveform profiles can be alternatively obtained by synthesizing the complex optical spectrum of the comb in a line-by-line manner with the aid of a pulse shaper [63]. This permits to engineer the optical intensity profile and allocate the available energy after photodetection into the spectral region of interest [118] [119].

An important aspect of optical frequency combs implemented using the techniques highlighted in Sects 2.2-2.4 is that their characteristics can be manipulated with external control signals. In the context of radio-frequency arbitrary waveform generation (RF-AWG) this opens a door to implement dynamic waveform reconfiguration at faster speeds than what is possible by simply changing the mask in a Fourier pulse shaper [120]. For example, in the technique reported in [120], two CW lasers are fed into a single opto-electronic frequency comb generator. This creates two identical combs around the corresponding laser wavelengths. If the CW laser frequencies are chosen to be sufficiently far from each other, the frequency combs can be independently synthesized using different locations of the Fourier mask in a line-by-line pulse shaper. Then, by switching the CW lasers in a complementary way, the current generated in the photodiode will effectively generate two different RF waveforms. A key point of this technique is that the switching is controlled by a pattern code whose temporal length is just limited by the unit memory. This in turn permits to achieve ultrabroad RF waveforms that may be non-repetitive in time and operate over a long record length. In [121], the RF spectrum of the synthesized signal using this technique is tailored with a resolution much finer than the spectral resolution of the shaper. The reader looking for a detailed discussion of the technique is referred to [122].

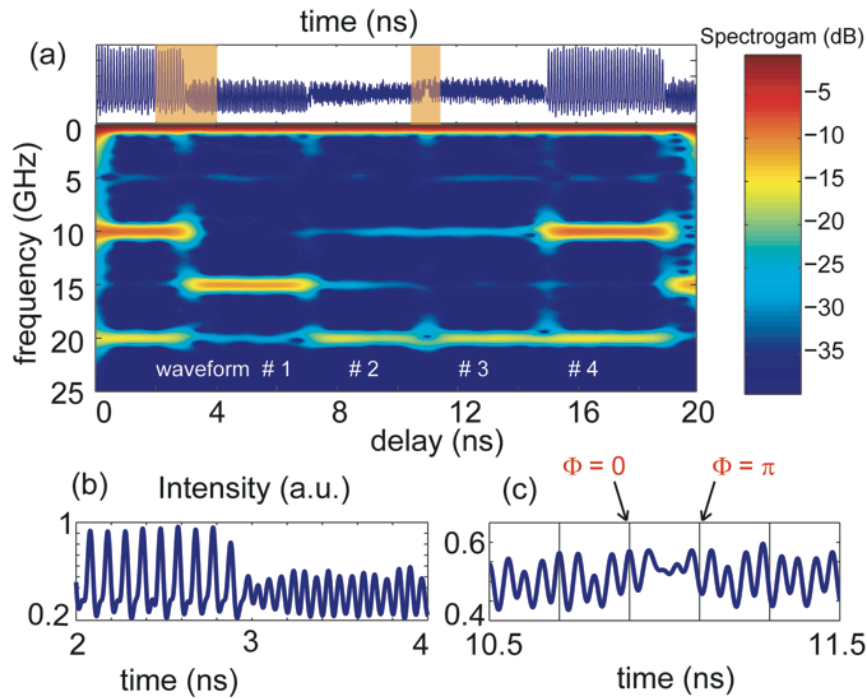

**Fig. 15**. *Multi-channel RF-AWG with rapid switching characteristics. (a) Single-shot intensity sequence showing transitions from 4 different synthesized waveforms, each lasting for 4-ns. The spectrogram is computed to help visualizing the frequency content of the different waveforms as they dynamically change over time. (b) and (c) are a zoom of the shadow regions in (a) where the transients between waveforms #4 to #1 and #2 to #3 occur, respectively. Adapted from* [126].

In a different technique [123], the switching is realized by shifting the central frequency of a single comb and interleaving two masks in a Fourier line-by-line pulse shaper with twice finer resolution. There are other approaches that effectively switch between independently synthesized RF waveforms, using, e.g., one comb and two independent shapers [124], or an array of lasers with predefined power and spectral separation [125].

The problem with the above techniques is that they are difficult to scale to include more than two independently synthesized RF waveforms. The challenge comes not only from the need for *N* switching elements but also from the need to get *N* independent high-speed control signals. In a more recent configuration [126], we showed that it is possible to switch *N* different frequency combs by realizing wavelength-to-time division multiplexing [127]. The reported scheme is highly scalable, since only one switching element and a single electrical control signal is needed for the *N* channels. The price to pay is the need for a faster switching signal and finer shaper resolution. Figure 15 shows an example using this approach. A sequence with 4 different RF-waveforms is synthesized with the aid of a 5-GHz resolution 2-D pulse shaper and 50-ps electrical control signals. The RF waveforms can then be effectively switched within 200-ps.

### 5.3 Ultra-broadband wireless coherent communications

In the previous subsection, we have seen that frequency-comb-based RF-AWGs combine broad-bandwidth characteristics with virtually infinite record length. As shall be discussed here, these are very appealing features for high-speed wireless communication systems.

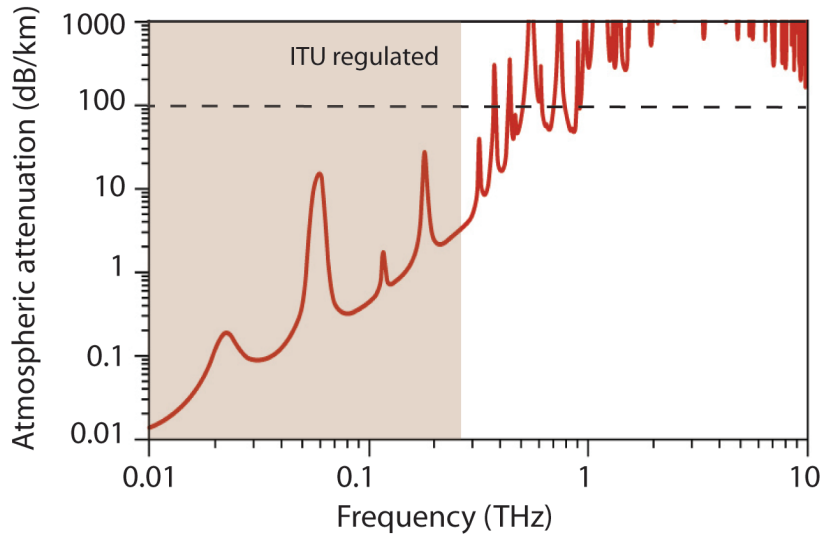

**Fig. 16**. *Atmospheric attenuation for radio frequencies at sea level and standard conditions (44% humidity and 20°C). The shadow region denotes the range that is regulated by the International Telecommunication Union (ITU). The dashed line denotes an attenuation of ~1dB every 10 meters. Adapted from* [130].

Today, several industrial and commercial applications demand higher data rates in wireless communication links. To cope with this thirst of bandwidth, the trend is to move technologies from current standards (such as WiFi or WiMax) that operate at low frequencies and limited bandwidth towards millimetre waves and even terahertz bands[1] in order to achieve rates approaching 100Gb/s [128].

However, as we move to higher frequency bands there appear new technological challenges and regulation constraints. At millimetre-wave frequencies and above, the atmospheric propagation provides natural windows around 90, 140, and 220 GHz [129], [130], as shown in Fig. 16, but different countries and organizations regulate the use of the electromagnetic spectrum in a different way. The 60GHz band (with ~10 GHz bandwidth) is an exception and is open for commercial purposes in most countries, but it shows a strong absorption peak. This restricts its practical use to indoor applications and a line-of-sight environment. Nevertheless, there are efforts to develop low-cost CMOS millimetre-wave circuitry, including antennas, amplifiers, oscillators, mixers, transceivers, and ADCs for > 10 Gb/s communications in this band [131].

---

[1] Strictly speaking, a millimeter-wave signal has a frequency carrier higher than 30 GHz. However, the convention in ultra-broadband wireless communication is to consider millimeter waves as those whose carrier frequency is higher than 40 GHz. Likewise, a signal earns the appellative "terahertz" starting from 300 GHz and not from 1000 GHz as one would a priori think.

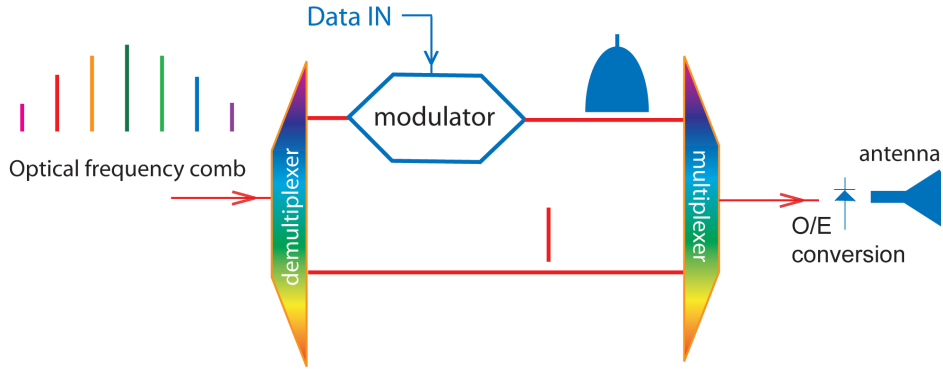

**Fig. 17**. *Frequency comb-enabled ultrabroadband RF photonic transmitter for high-speed wireless communications.*

The region beyond 275 GHz is not regulated yet, which means that in principle a larger bandwidth can be allocated for ultrahigh-data-rate wireless communications. The higher frequency limit is naturally imposed by the atmosphere attenuation around 500 GHz. The generation and wireless transmission of 10-100 Gb/s data signals within the terahertz band (300-500 GHz) is an active area of research where photonics technology plays a key role [132]. The synergy between high-speed wireless systems at millimetre-wave frequencies and fiber-optic communications is a topic that has been studied in the past and is probably better known by the name of "radio-over-fiber" [1]. Significant research efforts in the past have led to high-performance photonic/wireless components [133] such as high-speed uni-travelling carrier photodiodes [134], photoconductive antennas [135], and mode-locked laser diodes [136].

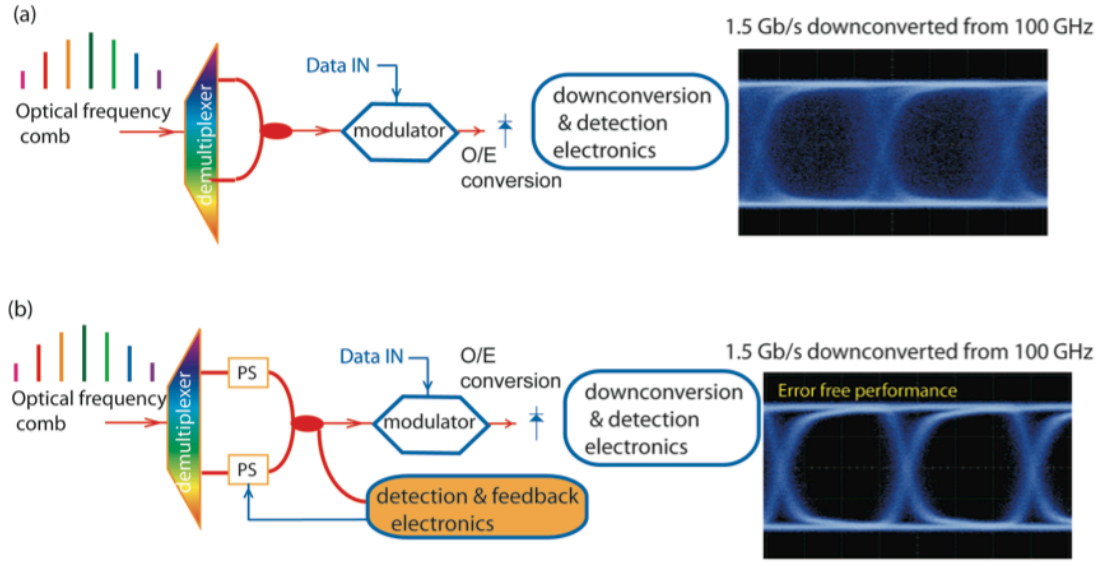

**Fig. 18**. *Frequency comb-based wireless transmitter with (a) fiber optic paths non-stabilized and (b) using a feedback mechanism to compensate for the fiber drifts using phase shifters (PS). The eye diagram data correspond to 1.5 Gb/s data encoded on a 100 GHz carrier that has been downconverted to baseband. Eye diagrams courtesy of Tadao Nagatsuma. Results adapted from* [137].

High-repetition-rate optical frequency comb is an enabling technology for ultra-broadband wireless communications. The essence is to leverage their broadband phase coherence to implement a hybrid photonic transmitter. The general layout of a comb-based photonic/wireless transmitter is shown in Fig. 17. Two separate lines from a high-repetition-rate optical frequency comb are isolated with, e.g., an arrayed waveguide grating demultiplexer. One of the comb lines is modulated by the data, usually encoded in an on-off keying (OOK) format. Then the two lines are combined and detected by a photodiode. If the photodiode has sufficient bandwidth, it will generate a modulated high-frequency carrier fixed by the spectral difference between the two comb lines. From the techniques presented in Sect. 2, it is clear that terahertz frequency carriers can be easily generated. This configuration has two appealing characteristics. First, the amplitude and phase of the photodetected RF signal can be independently modulated if an optical modulator that modulates the in-phase (I) and quadrature (Q) of the optical field is used (I/Q modulator). This permits to generate complex modulation formats using advanced optical modulation techniques and achieve higher data rates at a high-frequency carrier while covering a narrow spectral bandwidth. Second, the signal can be transported through an optical fiber before photodetection. This means that the transmitter and receiver can be placed at further locations than it would be possible if purely wireless transmission technologies were used.

Note that if the data input is encoded on a simple OOK format, the modulator can be placed after the multiplexer to modulate simultaneously both lines. But even in this case, keeping the phase stability of the two optical frequencies after recombination is a challenge using fiber-optic components. This issue was recently discussed in an experiment presented in [137] and the results are briefly summarized in Fig. 18. Two lines separated by 100 GHz coming from an optoelectronic frequency comb generator are isolated and combined using a fiber-optic coupler. After 1.5 Gb/s OOK modulation and subsequent detection in a high-speed photodiode, a 100GHz modulated wireless signal is generated. The signal is brought back to the RF baseband with a broadband mixer and analysed with a high-speed receiver. However, although the comb lines keep a stable phase relation, the eye diagram is not sufficiently clean to achieve an error free performance without applying forward error correction (FEC) techniques. The same group modified the setup of Fig. 18a to include an active stabilization scheme to correct for the phase drift induced by the fiber thermal instabilities that affect the phase of the high-frequency carrier. As illustrated in Fig. 18b, the eye diagram from the downconverted data becomes then clearly open, permitting to achieve error free performance, which is defined as a bit error rate (BER) < $10^{-10}$ without applying FEC.

The above simple example illustrates the relevance of preserving the comb's phase coherence after their individual manipulation. In the case in which the two lines can be simultaneously modulated, an integrated optical filter, resilient to environmental fluctuations, can actually be used to achieve a high-frequency carrier without the need for an active stabilization setup [138]. In the work reported in [139], 10Gb/s OOK at 94 GHz was successfully transmitted wirelessly over 800m. The two comb modes used to generate the millimetre-wave signal were filtered with an arrayed waveguide grating and summed together with an integrated combiner. However, it is more challenging to integrate active and passive components in the same chip to realize a configuration like the one presented in Fig. 17. Alternatively, recent advances in digital signal processing allow for implementing digital phase-recovery algorithms [140]. For example, in [141], 100 Gb/s photonic/wireless transmission at a frequency carrier of 87.5 GHz using two free running lasers and combining digital signal processing (DSP) with advanced modulation formats was successfully demonstrated. However, the achieved BER needs FEC in order to get error free performance, which was not implemented in the experiment.

## 5.4 Frequency-comb-based optical coherent communications

One of the most successful technologies in the history of fiber-optic communications is wavelength division multiplexing (WDM). The basic scheme of a WDM transmitter is shown in Fig. 19a. Several optical carriers located at different frequencies are independently modulated by external modulators with the data to be transmitted. The spacing between channels is standardized by the ITU, typically with 25, 50 or 100 GHz separation. The unit cell formed by an optical carrier in conjunction with its modulator and data sequence forms a single channel. On the receiver side, the information from a particular channel is then retrieved by filtering the corresponding band [142]. The success of this technology has been possible thanks to the advent of in-line broadband fiber amplifiers.

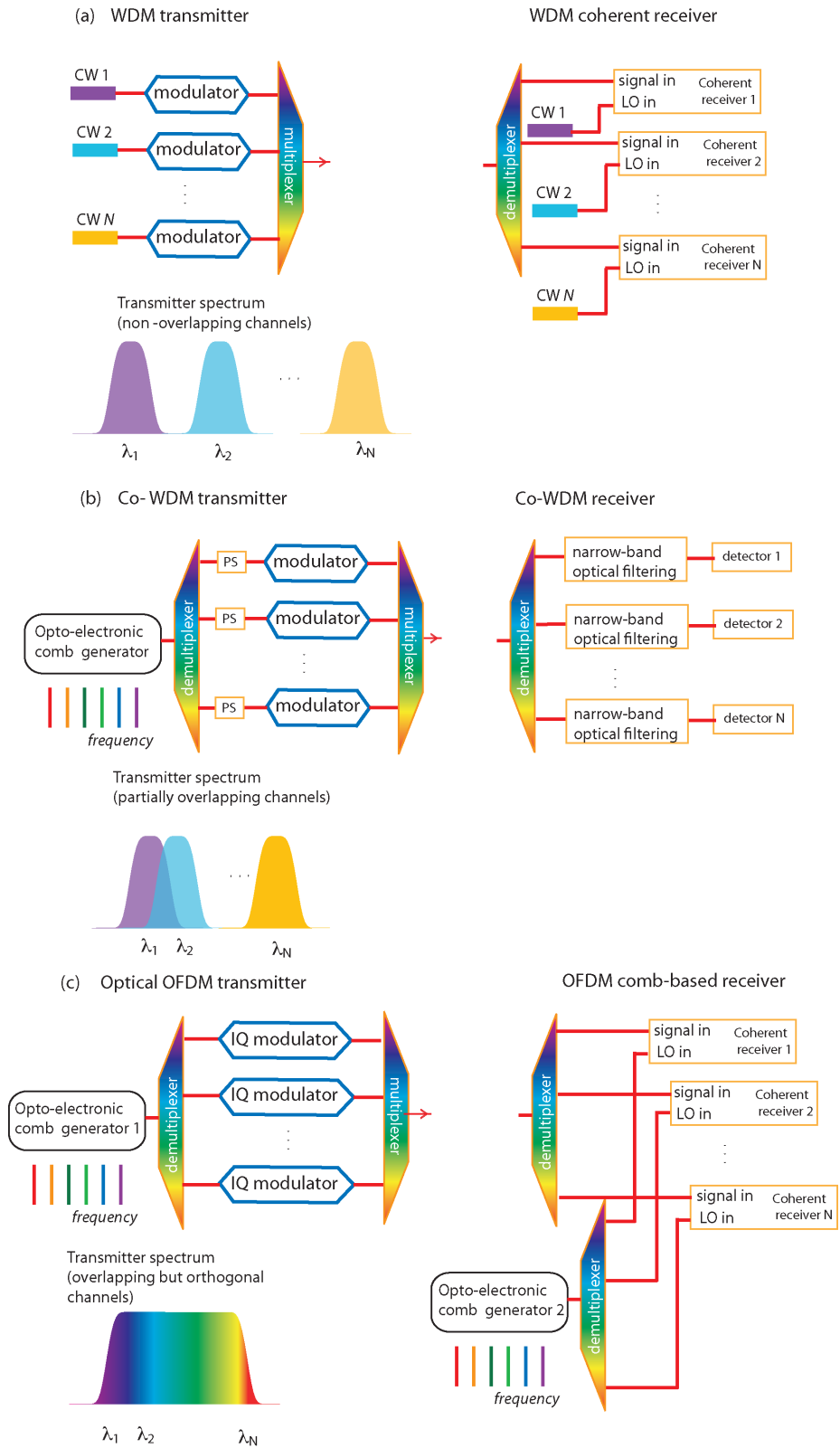

**Fig. 19**. *Frequency comb-enabled coherent optical communications. In (a) WDM transmitter the spacing between channels is >1.2x the symbol rate and therefore they can be individually filtered out. The modulator at each channel can be encoded using OOK or advanced modulation formats. For the latter case, at the receiver side, each channel must be measured in amplitude and phase, which is done with the aid of a coherent receiver. In (b) Coherent-WDM (Co-WDM) [150] and (c) Optical OFDM, the channel spacing equals the symbol rate, which results in partially overlapping spectra after subcarrier modulation. In Co-WDM (b), the receiver side realizes narrowband optical filtering (for simplicity shown for OOK). The introduced linear cross-talk is minimized by controlling the relative phase between lines at the transmitter with the aid of phase shifters (PS). In OFDM (c), the whole transmitted signal is detected, which may be done an optical frequency comb generator. This particular implementation of OFDM can be considered as a generalization of the coherent WDM transmitter in (a), where the different channels are measured and stitched together in the spectral domain.*

As an example, within the ~40 nm bandwidth available in the optical communications *C*-band, and considering 25 GHz frequency separation between carriers, ~ 200 channels can be easily allocated. With an OOK modulation format and at 10 Gb/s per channel, ~2 Tb/s can then be achieved in a single WDM transmitter. In this scenario, an optical frequency comb generator allows for replacing the large number of lasers required in the transmitter by a single one and fixing the channel separation across the bandwidth with great accuracy. In a 2006 experiment, more than 1000 channels were successfully generated from an electro-optic frequency comb generator working at 6.25 GHz repetition rate [143]. The frequency comb bandwidth spanned the *C*- (1530-1565 nm) and *L*- bands (1565-1625 nm) by using an external broadening technique, and >2.67 Tb/s transmission in an installed fiber network was successfully achieved by modulating the carriers at 2.67 Gb/s with OOK and applying FEC processing at the receiver.

The spectral efficiency of an optical communication transmitter is defined as the ratio between the bit rate and the optical bandwidth needed to achieve it. In WDM and for OOK, the spectral efficiency is < 1b/s/Hz for a single polarization. However, the achievable spectral efficiency in an optical fiber communication system is estimated from an information theory analysis to be at least ~ 6 b/s/Hz/polarization for a transmission distance of 2000 km [144]. This limit is fundamentally imposed by the noise of the amplifiers and the inherent nonlinearities and losses in single-mode fiber [145]. In order to reach high modulation efficiency, advanced modulation formats are needed that encode the digital information by exploiting different physical degrees of freedom available from the light source [146]. The reader looking for an introduction into the field of coherent communications is referred to [140], where different modulation formats, receiver architectures and digital signal processing techniques are introduced in a comprehensive way. Here, it will be sufficient to state that these modern modulation formats are indeed compatible with WDM [147]. As a matter of fact, the transmitter side would look like the one in Fig. 19(a) but with specialty modulators instead, capable of modulating all four quadratures of the optical field. On the receiver side, this certainly brings new challenges owing to the need to measure the electric-field variations at a multi-GHz rate (as opposed to only the intensity, as in OOK). For this aim, a homodyne scheme is adopted, where a well-calibrated narrow-linewidth laser is used as local oscillator (LO) for every channel. In modern coherent optical communication systems, the carrier phase is tracked with the aid of a high-speed digitizer and digital signal processing techniques [140]. A multi-channel WDM coherent receiver would look like the right hand side in Fig. 19(a). Here, an optoelectronic frequency comb generator might be used to replace the need for one LO laser per channel. In [148], an optical frequency comb generator was used as a coherent 4-wavelength LO to measure amplitude and phase variations of broad (40 GHz) spectral slices coming from an input arbitrary broadband waveform with long record length.

In order to improve the spectral efficiency, other research efforts aim at narrowing the spectral distance between channels. In multi-carrier systems, the optical bandwidth from a single channel is proportional to the symbol rate that drives the modulator. In the scheme of Fig. 19(a), each channel can be easily isolated without inter-symbol interference if the spacing is about 1.2x the symbol rate [149]. However, there are advanced coherent techniques in multi-carrier transmission systems, where the wavelength spacing equals the baud rate [149]. Two of these emerging examples are schematically presented in Fig. 19(b) and (c), respectively, i.e., coherent-WDM (Co-WDM)[2] [150] and optical orthogonal frequency division multiplexing (OFDM) [151]. In both techniques, optical frequency comb generators play a key role. Let us provide some details. After independent modulation and recombination, the signal complex field at the transmitter side, $e_T(t)$, both in Fig. 19(b) and (c), can be mathematically described as

$$e_T(t) = \sum_{k=1}^{N} \exp\left[-j\omega_k t\right] \exp\left[j\phi_k\right] m_k(t), \tag{2}$$

where $\omega_k$ represents the angular frequency of the *k*th optical subcarrier, $\phi_k$ a static phase, and $m_k(t) = \sum_{l=-\infty}^{\infty} a_{lk} m(t - lT)$ the modulation sequence, with $a_{lk}$ being, in the most general case, a complex symbol, $m(t)$ the pulse shape of the modulated line (assumed to be the same for the *N* subcarriers) and *T* the period. At the receiver side, the information carried by the subcarrier *p* can be recovered with the aid of a suitable optical band-pass filter with complex transmission spectrum $H(\omega - \omega_p)$. The demultiplexed electric field becomes

$$e_{R,p}(t) \propto \exp\left[-j\omega_p t\right] \sum_{k=1}^{N} \sum_{l=-\infty}^{\infty} a_{lk} \exp\left[j\phi_k\right] \exp\left[j(\omega_p - \omega_k)lT\right] a_{pk}(t - lT), \tag{3}$$

with

---

[2] It is important not to confuse the terms WDM coherent communications (whose architecture is displayed in Fig. 19a) with Coherent-WDM (Co-WDM, shown in Fig. 19b). The former one makes use of a coherent receiver per channel and sets a channel separation >1.2x the symbol rate. On the other hand, Co-WDM sets a channel spacing equal to the symbol rate and minimizes the incurred cross-talk by controlling the relative phases in the transmitter side.

$$a_{pk}(t) = \int \exp\left[-j\Omega t\right] M(\Omega + \omega_p - \omega_k) H(\Omega) d\Omega, \tag{4}$$

and $M(\Omega)$ being the Fourier transform of $m(t)$. Equations (3) and (4) indicate that in demultiplexing channel *p* there is a contribution from the neighbour ones. The impact of this cross-talk depends on both the relative phase of the subcarriers in the transmitter side and the shape of the filter.

One approach that benefits from a broadband and phase coherent multi-carrier source to minimize the cross-talk is Co-WDM [150] [152]. Here, the teeth of an optical comb provide the physical platform for the phase-locked carriers $\omega_k$, as illustrated in Fig. 19(b). The key point to note is that, owing to the phase-stable nature of the source, the cross-talk distortion introduced is deterministic (it is an interference term). In [152], the authors demonstrated that by fixing a constant π/2 phase difference between channels and optimizing the filter bandwidth at the receiver, the inter-channel distortion is minimized. This can be intuitively understood as follows. Assume the filter already substantially reduces the cross-talk fields. If the cross-talk field from one of the neighboring channels and the field from the desired channel are in phase, the variation in the received intensity is linear in the crosstalk field, i.e., we must add the *fields*, not the intensities. A crosstalk contribution after the filter equal to 1% in intensity corresponds to 10% in field. Summing the desired and crosstalk fields and squaring to get the total received intensity yields $\pm 20\%$ variation in received intensity. Similar considerations have been discussed, for example, in analyzing the effect of crosstalk from adjacent comb lines in optical arbitrary waveform generation via line-by-line pulse shaping [153]. But if the phases have a quadrature relationship, it is the *intensities* that add, and the power variation in the example above is only 1%, dramatically less than the in-phase case. This means that the cross-talk penalty induced by channels *p-1* and *p+1* on demultiplexing channel *p* affects only its perpendicular field quadrature. Figure 20 shows a recent experiment, where three different independently DPSK modulated comb lines were successfully received after narrowband filtering followed by balanced photodetection. Note that error free performance for all the three channels is achieved only when the phases of the channels are optimized.

It is important to note that to keep the quadrature phase relationship, we not only need stable carrier phases, but we also require that the carrier spacing exactly equals the symbol rate, i.e,

$$\omega_k - \omega_{k-1} = 2\pi/T \tag{5}$$

Although the relative phases of the fields originating from different carriers evolve in time, the carrier spacing specified above guarantees that the relative phases are periodic with the symbol period *T*. As a consequence, the level of cross-talk interference varies as a function of time within the symbol period. This effect is particularly clear in the eye diagram of the center line shown in Fig. 20. However, at the sampling instants the nearest neighbor crosstalk contributions are in quadrature with the center channel field, suppressing the overall intensity variation and yielding an open eye.

To date this approach has been only demonstrated for modulation formats like OOK [150] and DPSK [152]. However, in a recent theoretical study [154], it has been shown that higher-order modulation formats can also be implemented in Co-WDM provided that the signals driving the in-phase and quadrature branches of the modulator are delayed by half period.

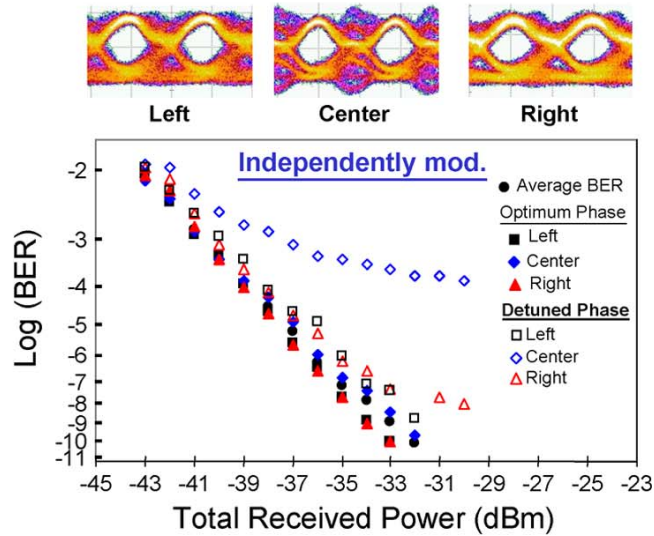

**Fig. 20**. *Frequency comb-enabled Co-WDM using DPSK. Three comb lines are independently modulated. The spacing between the channels equals the baud rate (32 Gb/s in this case). Error free performance for all channels is recovered when the phase of the comb lines is optimized. Adapted from [152].*

Other approaches to achieve higher spectral efficiency are based fundamentally on coherent (IQ) modulation and processing in conjunction with the spacing between channels again set exactly to an integer multiple of the symbol rate. From Eqs. (3) and (4), it can be shown that when eq. (5) holds and either the spectral filter $H(\Omega)$ is rectangular with $m(t)$ a sinc pulse [or $H(\Omega)$ is a sinc and $m(t)$ rectangular], then the information on the carrier $p$ can be extracted without linear cross-talk penalty. In this context eq. (5) constitutes an orthogonality condition. The first combination is at the heart of so-called Nyquist WDM [155], but further discussion of this topic is beyond the scope of this review. The second combination refers to the optical version of *orthogonal frequency division multiplexing* (OFDM), another modulation format that is gaining increasing popularity [151]. This multi-carrier modulation format was first proposed in the realm of wireless communications [156]. The main difference with respect to Co-WDM comes from the way the information is encoded and decoded. In mathematical terms, the starting point is Eq. (2). In OFDM, the different frequency modulated subcarriers have partially overlapping spectra and form a continuous spectrum when added together [see Fig. 19(c)]. From Eq. (2), we note that at a fixed sampling time, the transmitted signal resembles the inverse Fourier transform of an array of complex symbols [157]. Hence, upon transmission, the receiver may receive the information by inverting the process, i.e., by performing Fourier transformation of the transmitted OFDM signal. With the $H(\Omega)$ and $m(t)$ as specified and with the received signal sampled at a time properly synchronized to the OFDM symbol, the contributions from different carrier terms become orthogonal, leading ideally to zero crosstalk. This result may be traced to the well-known Fourier orthogonality condition, i.e.,

$$\frac{1}{T}\int_0^T e^{j\omega_k t}\, e^{-j\omega_l t} dt = \delta_{kl} \tag{6}$$

One of the main advantages of OFDM is that, for a fixed bandwidth, when you increase the number of channels, the bandwidth per channel goes down, providing a higher resilience to dispersion than other modulation formats. It is also interesting to note that unlike Co-WDM, OFDM is not sensitive to the choice of the relative phases of the carriers. Indeed, it is this property that permits immediate compatibility with high-level phase modulation formats. Thus, with respect to optical frequency combs, OFDM can take advantage of the equal carrier spacing property of such sources.

There are several ways to implement an optical OFDM signal, the main differences being the physical realization of the Fourier transformation operation and/or the optical subcarriers. The most popular approach consists of generating the subcarriers and complex coding in the electrical domain, followed by digital signal processing and digital-to-analog conversion (DAC). At the end of the DSP chain, two analog waveforms are generated and used to drive the two branches of an IQ optical modulator, which is fed by a single CW laser source [157]. This would generate the desired optical OFDM signal. Note that here, the subcarriers $\omega_k$ are physically implemented with the aid of the signal from the DAC. The main drawback of this approach is the substantial DSP required and the fact that the OFDM bandwidth will be ultimately limited by the DAC's bandwidth (just slightly above 10 GHz with present technology). This problem can be overcome by using an optical frequency comb as the multi-carrier generator [158] [159], as in the scheme shown in Fig. 19(c). Here, the comb lines play the role of the subcarriers $\omega_k$ and hence must be independently modulated in both amplitude and phase with an IQ modulator to provide the OFDM signal. This approach imposes another challenge in the receiver side. For this aim, another optical comb could be used as the multi-carrier local oscillator in a homodyne coherent receiver scheme [74]. Alternatively, as demonstrated in [159] [160], the receiver can implement the Fourier transformation all-optically in a cascade of delay interferometers followed by high-speed optical gating.

Note that in the Co-WDM and OFDM schemes depicted in Fig. 19, the subcarriers are wavelength multiplexed. This would help to reduce the losses with respect to the use of a coupler, thus ensuring the scalability of the approach. However, it would also bring new challenges owing to the need for correcting for the spectral response of the multiplexer [161]. Another challenge regarding the scalability of these two approaches is related to the need for keeping the phase coherence of the comb upon modulation and combination. We note that to date proof-of-principle demonstrations of Co-WDM and optical OFDM have generally been implemented using couplers, active feedback stabilization mechanisms, and just a few independently modulated channels.

## 6. Discussion

In the future, it is likely that optical frequency comb generators and pulse shapers will be developed in a compact platform to reduce their size, cost, weight, and power consumption, helping to enable applications outside the laboratory. In this direction, the use of photonics devices engineered at the nanoscale is considered as crucial for the next generation of optical communications [162] as well as photonic microwave signal processing [163] systems.

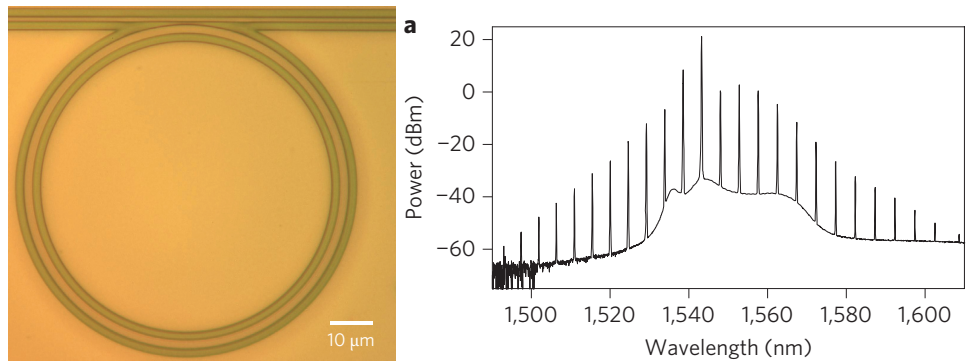

**Fig. 21**. *Example of a silicon nitride ring resonator (left) and emerging frequency comb generator when the ring is pumped with a CW laser closely matching a resonance. Results reproduced from [170] with permission from McMillan Publishers.*

Recent developments have led to an implementation of optoelectronic frequency comb generators using indium phosphide as the main platform [29]. This is a relatively mature technology with an extensive library of components, including lasers, amplifiers, and modulators. The comb generator presented in [29] consists of quantum well phase modulators that are embedded in a ring provisioned with gain via a semiconductor optical amplifer (SOA). The amplifying ring geometry allows light to accumulate several passes through the phase modulator before exiting the device, thus increasing the effective modulation index. This type of structure can be considered as an integrated version of the one presented in Sect. 2.1. When the parameters were correctly set, a phase-coherent multi-wavelength source that generates ~ 8 tones at -10 dB bandwidth with 10 GHz separation was achieved. The advantage of this approach is the capability to integrate further the structure with other laser sources while retaining the possibility to use an external RF synthesizer to set the comb spacing.

In 2007, a monolithic frequency comb generator based on a radically different scheme was reported [164]. It consists of a non-linear ring resonator with a loaded $Q$-factor ~$10^8$. The effective nonlinearity is enhanced by the long cavity lifetime and produces amplification of spontaneously generated photons through a four-wave-mixing process when the ring is pumped with a high-power CW laser whose wavelength is set to closely excite a resonance. In this scheme, the repetition rate of the comb is fixed by the FSR of the cavity (which can be easily fabricated to yield >100 GHz spacing). The scheme has been reported using different materials, e.g. $CaF_2$ [165], $MgF_2$ [166], and silica [164], in whispering gallery mode resonator structures, and with high-index glass [167] and silicon nitride [168] in planar chip geometries ($Q$~$10^6$ in the planar structures). The appeal of silicon nitride is its compatibility with CMOS industry fabrication standards and the possibility for further integration with other subsystems, such as multiplexers and modulators, onto the same chip. The ultrahigh-repetition rate, together with the compactness of this platform holds significant promise as a multi-wavelength light source not only for optical communications, but also for optical interconnections, astronomical spectrometers and broad-band RF photonics [169]. However, the above explanation for the comb generation process is heuristic. How exactly this process occurs and what are in turn the comb's spectral coherence properties is an ongoing topic of research [170] [171]. For the comb to be useful for the promising applications, it is of critical importance to answer the question under which conditions low-phase-noise performance can be achieved. What appears to be clear is that the generation process is independent of the material platform [172], yet extremely susceptible to the dispersion properties, which in turn control the non-linear phase-matching conditions. Exploratory experimental research on the implications of this platform in optical communications is just at its birth. An illustration of a typical high-Q silicon nitride ring resonator and the emerging comb spectrum is depicted in Fig. 21.

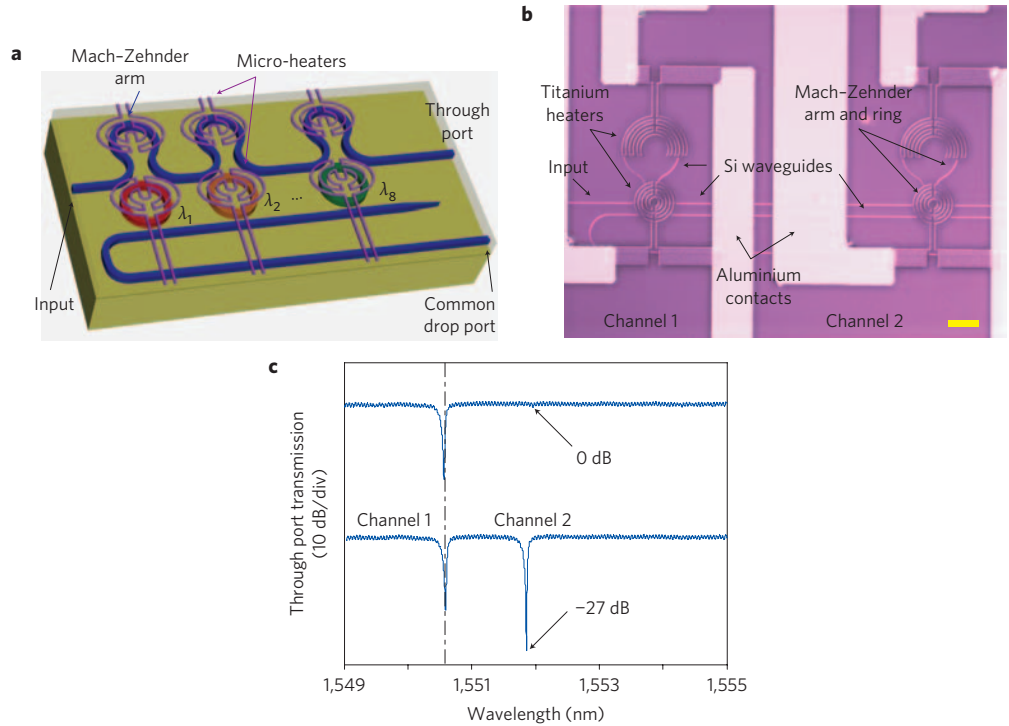

**Fig. 22**. *Example of an integrated pulse shaper fabricated in silicon-on-insulator technology. It has 8 independent attenuation channels. The transmission channels correspond to different ring resonators whose resonant wavelength and coupling coefficient can be independently controlled with the aid of two titanium heaters per ring. Results reproduced from* [174] *with permission from McMillan Publishers.*

There is a renewed interest in the development of silicon-on-insulator technology for the manufacture of photonic devices [173]. This technology is also compatible with CMOS fabrication industry standards and permits to achieve significantly smaller footprints than those available with silica lightwave circuits for passive devices, such as pulse shapers [174] or arrayed waveguide gratings [175]. In principle, this could enable the merging of frequency comb generators with more complex functionalities on the same platform. The work in [176] presented a 8-channel spectral amplitude carver with linear ring resonators, leading to a channel spacing of 1.8 nm and an out-of-band rejection ratio of ~ 40 dB. In the integrated pulse shaper in [174], developed for applications in photonic generation of arbitrary radio frequency waveforms, the channel allocation (resonant wavelength) and attenuation were independently controlled with a pair of titanium heaters per ring, one that sets the resonant shift and another the coupling ratio (and thus the loaded Q-factor). An image of the device and an example of single-channel attenuation are shown in Fig. 22.

Other recent efforts have focused on the development of passive delay lines [177] [178] and band-pass filters [179]. In the field of RF photonics, this type of devices has been used to demonstrate, e.g., narrowband RF photonic filtering [180] [181], downconversion [182], and optoelectronic oscillators [183]. A recent review on integrated microwave photonics has been presented in [163], covering the myriad of materials and technologies available. Although the complexity of the RF photonic filters demonstrated so far with integrated approaches in terms of flexibility and programmability is still far from the demonstrations presented in Sect. 4, this field is getting significant momentum. We expect to see in the future high-performance frequency comb generators and programmable devices with more subsystems all embedded on a single chip. Most likely, it will be the application's specific constraints that will dictate which is the most suitable technology platform. In this sense, efforts towards hybrid integration constitute a reasonable yet challenging direction.

## 7. Summary

The ideal optical frequency comb laser must be well conceived to target a particular application. In this work, we have provided an overview of the technology available to generate optical frequency combs at high repetition rates (>10 GHz) using standard fiber-optic communication equipment. We have placed a particular emphasis on opto-electronic frequency comb generators (typically implemented using electro-optic modulators). This type of platform is relatively simple to assemble and provides robustness, independent tuning of the central wavelength and repetition rate, and is therefore well suited for applications in ultra-broadband radio-frequency photonics. We have taken the broadest possible definition of this field i.e., considering any potential application offered by merging microwave engineering with state-of-the-art photonics technology. In particular, we have highlighted emerging frequency comb applications in optical communications, RF processing, arbitrary waveform generation, and THz wireless communications.

## Acknowledgments

We thank Magnus Karlsson for the careful reading of Sect. 5.4.

Victor Torres gratefully acknowledges support from the Swedish Research Council (VR) and the EU Seventh Framework program through a Career Integration Grant (Grant Agreement 618285). Andrew Weiner was supported in part by the Naval Postgraduate School under grant N00244-09-1-0068 under the National Security Science and Engineering Faculty Fellowship program and by the National Science Foundation under grant ECCS-1102110.